\documentclass[12pt]{article}

\newcommand{\beq}{\begin{eqnarray}}
\newcommand{\eeq}{\end{eqnarray}}
\newcommand{\beqs}{\begin{eqnarray*}}
\newcommand{\eeqs}{\end{eqnarray*}}

\usepackage{graphics,epsfig}
\usepackage{latexsym}
\usepackage{amsthm}
\usepackage{amssymb,amsfonts,amsmath}

\begin{document}

\title{The Strength of Regolith and Rubble Pile Asteroids\footnote{Accepted in Meteoritics and Planetary Science, 2/2014.}}
\author{P. S\'anchez and D.J. Scheeres \\ The University of Colorado at Boulder}
\date{}
\maketitle

\begin{abstract}

We explore the hypothesis that, due to small van der Waals forces between constituent grains, small rubble pile asteroids have a small but non-zero cohesive strength. The nature of this model predicts that the cohesive strength should be constant independent of asteroid size, which creates a scale dependence with relative strength increasing as size decreases. This model counters classical theory that rubble pile asteroids should behave as scale-independent cohesionless collections of rocks. We explore a simple model for asteroid strength that is based on these weak forces, validate it through granular mechanics simulations and comparisons with properties of lunar regolith, and then explore its implications and ability to explain and predict observed properties of small asteroids in the NEA and Main Belt populations, and in particular of asteroid 2008 TC3. One conclusion is that the population of rapidly rotating asteroids could consist of both distributions of smaller grains (i.e., rubble piles) and of monolithic boulders. 

\end{abstract}

\section{Introduction}

The strength and morphology of small asteroids in the solar system remains an open and fundamentally interesting scientific issue. The strength of a rubble pile body will control how fast it can rotate before shedding mass or disrupting, influence the process by which binary asteroids are created, and could have significance for the mitigation of hazardous near-Earth asteroids (NEA) should this be necessary in the future. The morphology of these bodies, including the size distribution of boulders and grains internal to the system, the macro-porosity of these bodies, and the shapes and spin states of these bodies, are important for understanding and interpreting spacecraft imaging of asteroids, for predicting the end-state evolution of these bodies, and for gaining insight into their formation circumstances. Despite these compelling issues and questions, real insight on the strength of rubble pile bodies and their morphology remains elusive. In this paper we provide a brief review of past and current models of these bodies, and offer theory, data interpretation and simulations that could shed light on these properties. 

A key effect that operates on small asteroids, and which provides a specific motivation and defining example of evolution for this paper, is the YORP effect \cite{rubincam_YORP}. Small asteroids subject to the YORP effect experience a net torque due to solar photons interacting with the asymmetric surfaces of these bodies \cite{pravec_harris_rotation, rossi_rotation}. If an asteroid is a rubble pile body, consisting of a size distribution of boulders, grains and fines, it is an inherently interesting question to understand how such a self-gravitating assemblage of bodies will react should it be spun to rates where significant internal stresses are present within the body \cite{sanchez_icarus}. If the rubble pile has strength, i.e., if there exist cohesive strength between its components as has been posited in earlier research \cite{scheeres_cohesion}, then it may even be possible for the body to spin beyond the point where centripetal accelerations exceed gravitational attraction. The interplay between morphology and strength will directly influence how a rubble pile body will respond and react to such extreme events. 

Information on the strength and morphology of small asteroids arises from three main sources. First and foremost are space missions to these bodies. In the class of small asteroids, the Hayabusa mission to asteroid Itokawa, with a mean radius of 162 meters, has provided unprecedented measurements of the total mass, shape, surface morphology and material properties of a small asteroid \cite{fujiwara_science}. The NEAR mission to asteroid Eros also provides crucial insight, although that asteroid is large enough (mean radius of 8.4 km) to fall into a class of bodies we do not specifically focus on in this paper, as similarly we do not consider the dwarf planets Vesta or Ceres. Second are ground-based radar measurements, which can provide a detailed shape and spin state of asteroids that come close enough to Earth to be imaged \cite{AIII_radar}. Radar observations provide important insight into the shape morphology of these bodies and can also unambiguously detect the existence of binary members. Third are ground-based photometric observations of asteroids, specifically the measurement of spin rates, spin states, and binarity from light curves and the inferred sizes from overall absolute magnitude \cite{rotation_database}. These photometric observations provide a large database of asteroid size and spin rates which can be used to place constraints on asteroid theories and to test specific hypotheses. Along with these more traditional observations, we should also include the recent observations of rotationally disrupted main belt asteroids \cite{jewitt_P2010_A, jewitt_P2013_A}, as these observations can provide insight into the constituent grain sizes found in rubble pile asteroids. 
In the current paper we also utilize information from sample return analysis of asteroid and lunar regolith as sources of information and constraints for the development of a theory for the strength of rubble pile asteroids. 

\begin{figure}[h!]
\begin{center}
\hspace*{-2.5cm}\includegraphics[scale=0.5]{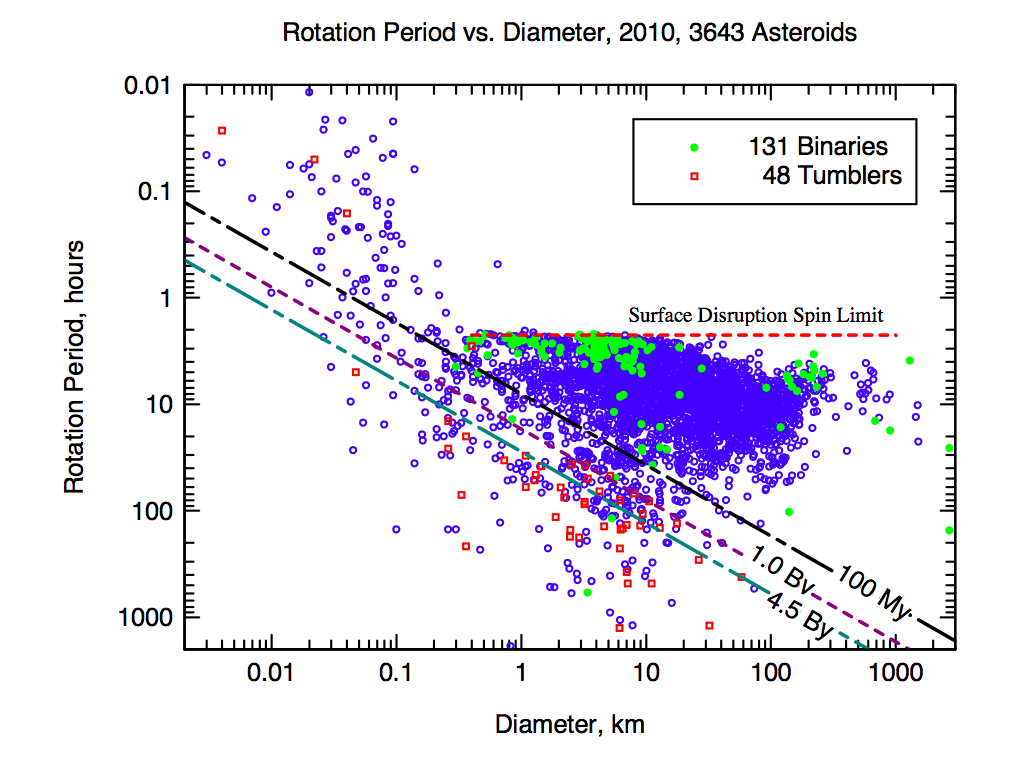}
\caption{Size/spin rate distribution of all asteroids, circa 2010. Binaries are called out in green, tumblers in red.
All of the spins and morphological types are taken from the asteroid lightcurve database \cite{rotation_database}. The slanting lines indicate relaxation timescales for tumbling asteroids, assuming material properties as specified in \cite{harris_relax}. }
\label{spinsize_harris}
\end{center}
\end{figure}

There exist current theories on the strength and morphology of small rubble pile asteroids that are largely motivated by a direct and simple interpretation of the asteroid size / spin rate curves. In their initial paper on this subject, Pravec and Harris \cite{pravec_harris_rotation} posited that all bodies spinning at rates beyond the surface disruption spin period of approximately 2.3 hours were coherent bodies, or monoliths. Figure \ref{spinsize_harris} shows this data with binaries and tumblers called out in different colors\footnote{A.W. Harris, personal communication}).  The surface disruption spin period is the spin period at which gravitational attraction is overcome by centripetal acceleration at the surface of a rapidly spinning spherical asteroid. According to this theory, rubble piles cannot be spun beyond this limit, as they will then undergo some sort of disaggregation into their component boulders, gravels and fines. This theory presupposes that rubble piles have no cohesive strength between their components and that the disaggregation process is eventually catastrophic and separates all of the constituent components of the asteroid. The remaining components are then small relative to the larger rubble pile bodies and can be spun by the YORP effect to spin rates that are much faster, as their material strength can sustain large tensile loads. 

In a series of papers, Holsapple \cite{holsapple_original, holsapple_plastic, holsapple_spinlimits, holsapple_yorp} pushes at several aspects of this theory to probe the relation between the deformation of a rubble pile body and its spin rate, and the level of cohesive strength necessary to keep fast spinning rubble pile bodies bound together. His analysis shows that the level of strength necessary to keep all of the known fast spinning asteroids bound together is modest at best, and corresponds to relatively weak rock \cite{holsapple_spinlimits}. His analysis of stress and failure within rapidly spinning rubble pile bodies also shows that the relationship between the observed spin period barrier and the deformation of a rubble pile is more complex. Specifically, he shows that when additional angular momentum is added to a rapidly spinning cohesionless rubble pile it may actually undergo a decrease in its spin rate (increase in spin period) due to body deformation. Granular mechanics simulations by S\'anchez and Scheeres \cite{sanchez_icarus} also show this behavior for cohesionless rubble pile bodies.  Holsapple doesn't specifically propose a theory for how cohesion could arise within rubble pile bodies or provide specific predictions of body morphology. However, his work introduces the application of continuum mechanics properties such as friction angle, cohesion, and failure theories to the realm of rubble pile bodies. It should also be mentioned that Sharma has been independently investigating the geophysics of rubble pile bodies using a similar continuum mechanics approach \cite{sharma_DP, sharma_structure}. 

Theories for the morphology of small rubble pile asteroids have been proposed in the past, although such models are difficult to prove without sub-surface knowledge of a rubble pile body. Britt and Consolmagno \cite{AIII_density} provide a heuristic description of the possible distribution of material within a rubble pile body in order to explain the observed high porosities of these asteroids. In their model they suppose that the interior of a rubble pile consists mainly of larger boulders with the finer gravels trapped, in some sense, at the surface of the asteroid. Other descriptive models have been proposed for the migration of grains within rubble pile bodies, specifically in \cite{asphaug_swift} an application of the ``Brazil nut effect'' is applied to a simple model of an asteroid surface. Along these lines, an analysis Itokawa images is made by Miyamoto et al. \cite{miyamoto_science} to show that there is flow of smaller gravels across the surface of Itokawa, with the finest material settling into the geopotential lows on the surface of that body. While simulations of asteroid disruption and reaccretion have been carried out, such as by Michel et al. \cite{michel2001collisions, michel2003disruption}, these computations generally do not resolve the dynamics and distributions of the meter size and smaller boulders and gravels that are created in such disruptive events. In the current paper we attempt to link a model for the interior morphology of a rubble pile body to its observed strength characteristics, motivated by a desire to establish what the observable characteristics of a body's interior morphology may be. 

The outline of this paper is as follows. First, we motivate our understanding of asteroids as rubble piles through a discussion of the size distributions of the asteroid Itokawa as inferred from the Hayabusa mission and through recent observations of disrupting main-belt asteroids. Following this, we develop a simple analytical model for the strength of a rubble pile asteroid described by a size distribution, using standard physics models of van der Waals cohesive attraction between grains. This motivates a view of rubble pile asteroids where interstitial grains of finer size may play an important role in holding larger boulders and grains together, enabling them to spin more rapidly. Following this we evaluate and validate this model of finer regolith holding larger grains in a matrix through a series of simulations that probe a more realistic model of how this effect can work. We then develop a simple model for an asteroid's strength to provide a theoretical link between cohesive strength and failure. This provides a bridge between our strength computations and predictions for how rapidly a rubble pile body could rotate before it undergoes failure -- either deformation or fission. Finally, we end the paper by explicitly comparing the theoretical limits defined by the theory with observational data on asteroids. 

\section{Rubble Pile Component Size Distributions}

This section reviews results from observations of the surface of Itokawa and analysis of its sample return. We also discuss properties of lunar regolith as a possible model for granular materials on an airless body. Finally we note recent observations of disrupted asteroids in the main belt, and comment on how they support aspects of our size distribution and asteroid morphology model. 

\subsection{Measured Size Distributions on Itokawa}

The Hayabusa mission to asteroid Itokawa measured several key morphological properties of that asteroid. 
Michikami et al. \cite{michikami} estimated the surface size distribution of rocks and boulders from millimeter to tens of meters in size and determined that it follows a cumulative distribution on the order of $d^{-2.8}$, where $d$ is the diameter of a grain. The largest boulder found on the surface was Yoshinodai, approximately 40 meters in size. They speculated that the lower limit for this size distribution was on the order of mm to cm, but this was not based on direct measurement. More recently, Mazrouei et al. \cite{mazrouei_itokawa} studied Itokawa imagery and determined a block size distribution up to $d^{-3.5}$ for blocks larger than 6 meter. Based on this range of measured distributions we will use an analytically tractable $d^{-3}$ size distribution (see the Appendix for more details on the $d^{-3}$ size distribution). 

The main purpose of the Hayabusa mission was to collect samples of the asteroid surface, which it successfully did \cite{itokawa_sampling}. The sampling mechanism that was flown on Hayabusa was supposed to fire a pellet into the surface to create a cratering event and collect the resulting ejecta field in a sample horn, which connected to a sample chamber \cite{yano_science}. Instead, the pellet did not fire during contact between the sample horn and the asteroid surface, although the spacecraft did collect a minute amount of material from the surface of the asteroid. It is significant that the sampling took place non-destructively, without fragmenting the asteroid surface with a high-speed bullet \cite{tsuchiyama_science}, meaning that the collected sample represents a sample of naturally occurring grain sizes on the asteroid surface. 
Analysis of the grains removed from the sample chamber yielded a size distribution of order $d^{-2.8}$ from $\sim 100$ microns down to 1 micron sizes, with the original sample potentially being as shallow as $d^{-2}$ prior to removal from the chamber due to disaggregation of larger clumps of grains into smaller constitutive pieces during removal. Tsuchiyama et al. \cite{tsuchiyama_science} speculate that this size distribution would extend up to the millimeter size range. We note that our theory is based on the size distribution of individual grains, and thus the $d^{-2.8}$ is the appropriate distribution for us to reference, similar to the macroscopic distribution. 

As a caveat, we note that both the macroscopic and microscopic size distribution measurements are based on surface observations and sampling, and do not provide any direct insight on the interior of the body. By using these surface derived size distributions for interior size distributions we make a clear assumption. Also, although both the macro observations and sample analysis indicate a similar size distribution, this does not prove that the true particle size distribution on this body should extend continuously from decameter-sized boulders down to micron-sized grains. This represents a second key assumption in this paper. 

Given the measured mass of Itokawa of $3.58 \pm 0.18 \times 10^{10}$ kg \cite{abe_science} and the revised volume of $1.77 \times 10^7$ m$^3$ \cite{gaskell_lpsc}, a revised bulk density of $2.0$ g/cm$^3$ can be computed, slightly different than that computed in \cite{fujiwara_science, abe_science}. Assuming a grain density of $3.4$ g/cm$^3$, as measured from the samples, yields a bulk porosity of 40.6\%, or packing fraction of 0.59. 

\subsection{Observed Grain Sizes from Disrupted Asteroids}

In a series of recent papers, Jewitt and collaborators have reported extensive observations of disrupted main belt asteroids, which they suppose have been disrupted through rapid spin rates \cite{jewitt_P2010_A, jewitt_P2013_A}. By tracking the evolution of the debris fields over time, Jewitt et al.\ place constraints on the size, number and volume of grains liberated from the asteroids. For the body P/2010 A2, observations of the debris field occurred at least a year after the event, and the observed grain sizes only extended down to the $\sim 0.1$ mm level \cite{jewitt_P2010_A}. However, in the more recent observation  of P/2013 P5, where the debris fields have been imaged just weeks after disruption, the observed grain sizes extend down to at least the 10 micron level \cite{jewitt_P2013_A}. We interpret these observations as further evidence that there exist substantial finer grains associated with rubble pile asteroids. While the detailed mechanics of how these bodies disaggregated are not available, based on detailed simulations of spin-disruption \cite{sanchez_icarus} we assume that the debris field contains material from the surface and at least the shallow sub-surface of the asteroids. 

\subsection{Synthesized Size Distribution and Packing}

\label{sec:2.3} 

If we assume a $d^{-3}$ distribution across the full span from micron to tens of meters we can explore a range of computed quantities for a rubble pile body. The Appendix carries out a series of computations related to this size distribution, some of which we highlight in the following. For our discussion we will be assuming Itokawa-motivated examples, with $N_1$ largest boulders of mean radius $r_1 = 20$ meters and smallest particles of $r_0 = 1\rightarrow 10$ micron (based on the Itokawa sample \cite{tsuchiyama_science}), and explicitly assume that $r_0 \ll r_1$. For such a size distribution we find that the mean grain size is $1.5 r_0$,  that the total surface area across the entire rubble pile is $12\pi N_1 r_1^3 / r_0$, and that the total volume is $4\pi N_1 r_1^3 \ln (r_1/r_0)$. Thus, we note that such a rubble pile will be dominated in number by the smallest grains, and that they will also dominate the surface area open for contact between grains. The volume, on the other hand, is distributed equally throughout the logarithm of grain sizes. A few comparisons make this point. The grain size at which half the total surface area is evenly split between  smaller and larger grains is the harmonic mean of the minimum and maximum grain sizes, $2r_0 r_1 / (r_1+r_0) \sim 2 r_0$, which means that the smallest grains in the distribution dominate the surface area distribution. The grain size at which half the total volume is evenly split between smaller and larger grains is the geometric mean $\sqrt{r_0 r_1}$, and thus for our above example would equal 4.5 millimeters. 

To evaluate whether there are sufficient quantities of small grains to coat the larger boulders we can carry out a few estimates based on the size distribution. For a boulder of radius $R$, the volume of material required to coat it to a depth of $\Delta R$ equals $\Delta V = 4\pi R^2 \Delta R$. Let us assume that we cover the surface with smaller spheres of radius $\Delta R$. Since the volume of a sphere will cover approximately 1/2 of the ``cube'' it can fit into, we make the assumption that if the boulder is covered with spheres of radius $\Delta R$ and volume $V_o = 4\pi/3 \Delta R^3$, that their total volume is $\sim \Delta V$. The number of grains required is found as $N_1 = \Delta V / V_o = 3 (R/\Delta R)^2$. Assuming our size distribution the number of grains available for one boulder at a size $R$ at a grain radius of $\Delta R$ is approximated as $N_o \sim n(\Delta R) \Delta R = 3 (R/\Delta R)^3$. Thus, the number of layers that the boulder can be coated with equals $N_o / N_1 = R / \Delta R$. As we will see, our model only requires a few layers of interstitial regolith to surround a boulder in order for the physics to work. Thus, for decameter-sized boulders and grains down to the 1-10 micron range we see that there should be more than enough fine regolith to cover the bodies. In fact, it appears that a more shallow size distribution should also be able to supply adequate covering, however we do not probe how shallow the size distribution can be.
This implies that there are ample grains to multiply cover larger boulders, at least partially filling the interstitial voids between them. 

\section{Regolith Strength Models and Observations}

In the asteroid environment the forces affecting a small asteroid composed of a size distribution of rocks, such as Itokawa, have traditionally been identified with gravity, internal friction, and inertial forces from rotation. These are all scale independent, and this fact has driven the study of rubble pile bodies for the last few decades. More recently, however, the work by Holsapple \cite{holsapple_spinlimits} has shown that some cohesion within an asteroid can strengthen it against spin disruption at rapid rotation. In Scheeres et al.\  \cite{scheeres_cohesion} it was shown that the expected strength of cohesive forces will rival and exceed these other forces as the size of a body is reduced, and thus must be accounted for in the study of their mechanics. This section reviews the measured strength of asteroid regolith and develops a simple analytical model for how regolith strength can vary with grain size. These results provide some foundation for our simulations in the following section. 

\subsection{Cohesive Forces}

Minerals exhibit weak attractive forces between collections due to several physical effects which are generally lumped together with the catch-all term van der Waals forces. In terrestrial settings, these forces and their effects are only significant at particle sizes below 100 microns, and become dominant for powders of size less than 10 microns \cite{castellanos}. The attractive force between two spheres of radius $r_1$ and $r_2$ in contact is approximated by 
\beq
	F_c & = & A_h r_1 r_2 / (r_1+r_2) \label{eq:vdw}
\eeq 
where $A_h$ is a material constant related to the Hamaker constant. For lunar regolith with an assumed ``clean'' surface the published values of $A_h$ are $\sim 0.036$ N/m \cite{perko} and we will use this as a representative value throughout the paper. For powders on Earth the significance of these cohesive forces only become evident when they become larger than the other forces in the system, most generally when cohesive forces are a factor of 10 or larger than the particle weight. At and below these sizes the mechanical properties of cohesive powders change significantly and they act as weak solids \cite{castellanos}.

Under this cohesion model the attraction between equal sized grains of radius $r$ is $\frac{1}{2} A_h r$. For an attraction between a grain of radius $r$ and a larger object of radius $R>r$ the net cohesive force is always greater than the attraction between equal sized grains. Thus, the model predicts that fine grains will preferentially attach to larger grains, and thus that larger grains embedded in a matrix of fine grains could be held in place by the strength of the matrix. This is a classical result in granular mechanics, and explains why cohesive grains preferentially coat larger intruder surfaces.This result also motivates our model of asteroid rubble pile strength, with larger boulders and grains being held in place by finer grains.  

\subsection{Computed Strength}

The strength that van der Waals cohesion between grains produces will be a function of how the particles are packed together. This can be computed, in principle, using the following standard approach. Let $A_b= r_b^2$ be the area of a cross-sectional cut, $r_b$ the dimension of this cut, $\phi$ the packing fraction and $C_\#$ the coordination number (the average number of neighboring particles that touch a given grain).  Then, the number of contacts would ideally be the product of the number of particles that are cut  and the number of contacts these particles have:
\begin{equation}
N_c=\frac{r_b^2 \phi}{\bar r_p^2}\frac{C_\#}{4}
\label{ncontacts}
\end{equation}
where $C_\#$ is divided into 4 as in average half of the particles in contact are going to be above the cut and the other half below it and only half of them would be in a contact angle that put them completely below (or above) the cutting surface.
Equation \ref{eq:vdw} is how we calculate the cohesive force between the particles, so for two particles of the same size (or the average size), this would be:
\begin{equation}
f_c=A_h \frac{\bar r_p}{2} \label{fvdw}
\end{equation}
where ${\bar r_p}$ is the average grain size. 
Equations \ref{ncontacts} and \ref{fvdw} give us the total number of contacts across the cross-sectional surface in a specified direction and the force per contact. Later we will determine correction factors to account for the randomized orientations of contacts, the existence of both tensile and compressive contacts, and further considerations. These corrections will be determined from our randomly packed simulations.
However, given our current ideal formulation the total force across the surface area for yield ($\sigma_{Y}$) would then be the result of putting together all these assumptions, which results in:
\begin{eqnarray}
\sigma_{Y} & = &\frac{N_c f_c}{A_b}\\
 & =&\frac{1}{r_b^2}\frac{r_b^2 \phi}{\bar r_p^2}\frac{C_\#}{4}A_h \frac{\bar r_p}{2}\\
& =& 0.125 A_hC_{\#} \phi \frac{1}{\bar r_p}
\end{eqnarray}
Let
\begin{equation}
s_{Y}= 0.125 A_hC_{\#} \phi 
\end{equation}
then $\sigma_{Y}$ has the form:
\begin{equation}
\sigma_{Y}=\frac{s_{Y}}{\bar r_p}
\end{equation}
For {\it C}$_{\#}$=4.5, $\phi=0.55$ (which is consistent with cubic packed regolith grains) and {\it A}$_h$=$3.6\times 10^{-2}$ N m$^{-1}$ (from lunar regolith measured values) we find the total strength to be
\begin{equation}
\sigma_{Y}=\frac{0.011}{\bar r_p}
\label{numeric-sigma}
\end{equation}
expressed in units of Pascals when ${\bar r_p}$ is given in meters. 

Under this idealized packing relation we find that a matrix of micron radius grains in a cubic crystal packing would provide over 11 kPa of strength, with a matrix of 100 micron grains providing over 100 Pa. Later we will see that when the grains are randomly packed we get a significant decrease of strength from this packing, and thus that these current computations are idealized and must be corrected for geophysical applications. 

\subsection{Properties of Lunar Regolith}

It is relevant to review the measured properties of lunar regolith, which are perhaps the most similar materials to asteroid regolith studied in the past, having similar mineralogical properties for some asteroid types and being generated by impact processes. We do note, however, that there may be substantial differences between lunar and asteroidal regoliths in terms of history and processing. Despite this, the upper uncompacted surface of lunar regolith (sometimes called a ``fairy-castle'' structure) could mimic the porosity and mechanical properties of asteroid regolith settled on and within the micro-gravity environment of an asteroid. A detailed description of lunar regolith properties determined from Apollo and terrestrial experiments is reported in \cite{mitchell_lunar} and summarized in \cite{colwell_lunar, perko}. 

The upper $\sim 15$ cm of the lunar regolith was observed to have a porosity similar to the implied bulk porosity of Itokawa, and similar to the simulated porosities in our experiment. 
Thus we take the properties of this well-studied granular material to be a possible analog to the distributions we discuss here. 
Specific measurements from the Apollo Soil Mechanics Experiment S-200 \cite{mitchell_lunar} have the following measured or inferred mechanical properties for the upper layer of lunar regolith:  porosity of $\sim 50$\%, internal friction angle of $\sim 40^\circ$ with variations on the order of 10$^\circ$, and cohesion ranging from $\sim 0.1$ kPa up to a few kPa. Deeper in the lunar regolith, where particles are more strongly compacted, porosity shrinks to 42\% (packing fraction of 0.58), the friction angle trends larger to over $50^\circ$, and the cohesion increases to $\sim 3$ kPa. 

\begin{figure}[h!]
\begin{center}
\includegraphics[scale=0.39]{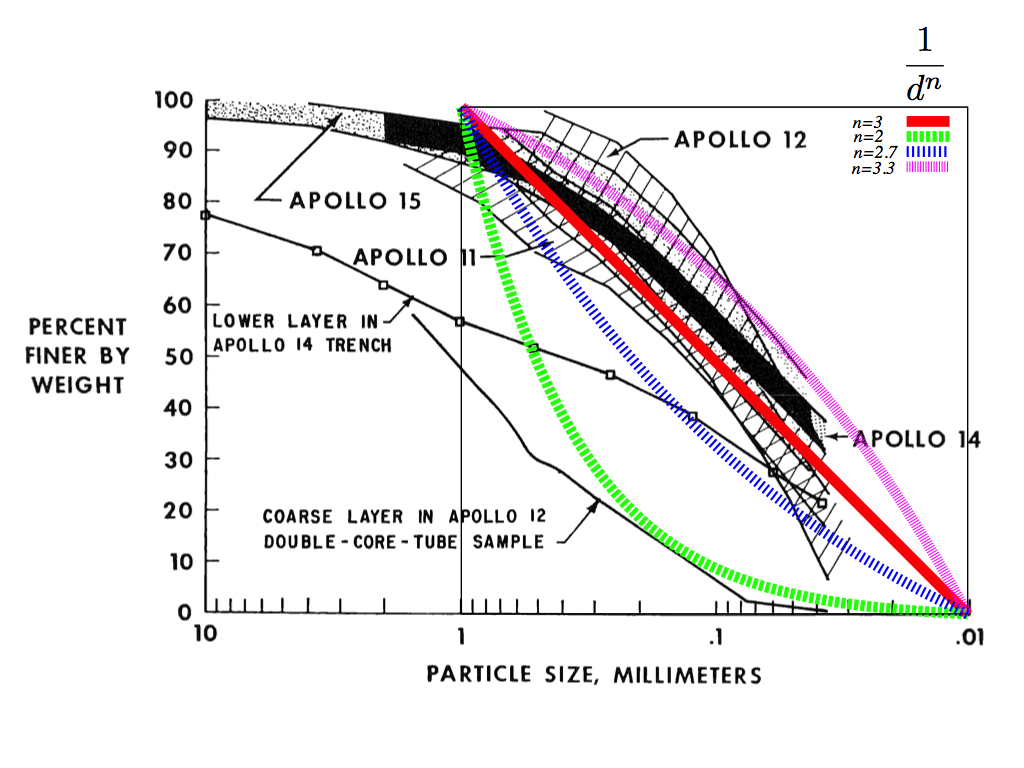}
\caption{Restatement of Fig.\ 1 from \cite{mitchell_lunar} with overlays of the ideal weight fraction for size distributions of $1/d^3$, $1/d^2$, $1/d^{2.7}$ and $1/d^{3.3}$ with smallest grain size of 0.01 millimeters. }
\label{fig:mitchell}
\end{center}
\end{figure}

The size distribution of grains in the lunar regolith are not explicitly presented, although in \cite{mitchell_lunar} a plot of weight less than a given grain size as a function of grain size is presented. The grain sizes are tracked from a few centimeters down to a size of a few tens of microns (see Fig.\ \ref{fig:mitchell}). Overlaid on the figure are a few ideal weight fraction curves for size distributions of $1/d^3$, $1/d^2$, $1/d^{2.7}$ and $1/d^{3.3}$ ranging from 1 millimeter to 0.01 millimeters. From a direct comparison we can conclude a size distribution similar to but slightly steeper than an ideal $1/d^3$ with a minimum grain size less than $\sim 10$ microns. From these comparisons we infer that using a $1/d^3$ grain size distribution is a reasonable approximation. 

\section{Simulating the Strength of an Asteroid}

The above computations are ideal, assuming a consistent structure for grain distribution throughout the body. The assumption is also being made that larger boulders will be ``held in place'' by a surrounding matrix of finer material. Both of these assumptions can be tested with simulation, allowing for the more realistic mechanics of randomly packed and settled polydisperse grains in a relevant gravitational environment. In this section we explore these assumptions by directly simulating the hypothesized situation with a direct soft-sphere discrete element model \cite{sanchez_ApJ, cundall1971}. 
Specifically, we settle smaller cohesive grains between two large gravitationally attracting boulders and pull on the boulders to find the yield strength of the assemblage. 
We note that these computations present yield strength computations for regolith that has had no forced packing and has been allowed to settle under self gravitational attraction alone, which will conservatively model accretion in a microgravity regime. 

\subsection{Simulation Description}

Our simulations model a self-gravitating system that consists of two spherical ``boulders'', 1m in size that, though not in contact, are connected to one another through thousands of smaller, cohesive, spherical particles of sizes ranging from 1 to 5 cm (see Fig.\ \ref{snapshots}).    This is what we have termed a ``granular bridge,'' as the granular counter-part of a liquid bridge.  In the latter, the specifics of the molecules of the liquid are overlooked in the benefit of the calculation of a net cohesive force among other properties.  Given that real size simulations cannot take into account the billions of individual particles, this is exactly what we will do.  We will develop a general model that can be compared to the theory and applied to the simulation of real size asteroids with a number of particles that is computationally feasible.

The entire system is contained in a rectangular box of $1.5\times1.5\times2.5$ m with periodic boundary conditions that only affect the contact forces.  
Our numerical code uses a Soft-Sphere DEM \cite{cundall1971, sanchez_ApJ} that implements the cohesive forces between spheres as a contact force following Eqn.\ \ref{eq:vdw}.  Normal contact forces are modeled through a linear spring-dashpot \cite{allen, herr1} and tangential forces (static and dynamic friction) are modeled as a stick-slip interaction through a linear spring that exerts a maximum force that respects the Coulomb yield criterion \cite{silbert}.  The difference between the spring-dashpot (repulsive) contact force and the cohesive Van der Waals (attractive) force produces a net interaction that will determine whether a contact is in tension or in compression. Self-gravitating forces among the particles forming the bridges are calculated following \cite{sanchez_ApJ, sanchez_icarus}.  For boulder-boulder, regolith-regolith and boulder-regolith gravitational attraction these forces are calculated exactly, considering all bodies as point masses.
These particles and the boulders have a grain density of 3200 kg m$^{-3}$.

Initially, the particles forming the regolith are placed in horizontal layers and in a hexagonal closed-packed (HCP) lattice; the distance from center to center is 1.1 times the size of the largest particle.  One boulder is placed above the layers and the other below them; this geometry helps us to produce very symmetrical systems.  The two boulders never touch one another.
The number of particles used in the simulations depends on the size of the particles so it is easier to refer to the systems by the number of layers that were formed (L1$_x$, L2$_x$, L3$_x$ and L4$_x$); the letter $x$ will be substituted by the size (monodisperse) or range of sizes (polydisperse) of the particles in centimeters.  When the particles were not monodisperse, they followed a $1/d$ size distribution with individual grains selected randomly from the stated limits \cite{recipes}.  More realistic size distributions are computationally challenging, with the number of additional smaller grains growing too large for conventional computational approaches. For example, for a $1/d^3$ size distribution ranging from 1 to 10 millimeters, for every 10 millimeter-sized grain we must introduce 1000 additional grains of smaller size, down to millimeter-sized. Computation of such steep size distributions is what we are trying to avoid by developing a model for how small regolith grains can stabilize larger boulders. 

Systems L1$_2$, L1$_3$, L1$_4$, L1$_5$ consisted of 5265, 2340, 1307 and 822 small grains of size 2, 3, 4, and 5 cm, respectively.  Systems L1$_{2-3}$ and L1$_{3-4}$ were formed by 2340 and 1307 small grains of sizes between 2 and 3 cm and 3 and 4 cm, respectively.  More layers only used multiples of these number.  
We can compare these layering numbers in terms of our earlier computations on the number grains required to cover a boulder with multiple layers. In Section \ref{sec:2.3} we found that a boulder of radius $R$ covered in grains of radius $\Delta R$ would in general have $R/\Delta R$ layers for a $1/d^3$ size distribution. For our system we find that our 1 meter boulders could be covered by 50 layers of 2 cm grains, and 20 layers of 5 cm grains. Since we only consider up to 4 layers of interstitial grains in our simulations, we see that our simulations are quite consistent with our modeling assumptions. 

In order to form the granular bridges, the particles in the regolith are given small random initial velocities and the boulders are allowed to move only in 1D and cannot rotate.  Applying the settling method detailed in Sanchez et al. 2004 and Sanchez and Scheeres (2010) \cite{sanchez_2004,sanchez_ApJ} we produce reproducible granular systems with extremely low kinetic energies ($E_r<10^{-6}$).  
Using these preparations, we then carry out our strength simulations. The procedure we use is followed for each case. 
No pull is applied during the first 5 seconds of simulation after the settling process finished.  At t=5 s we pull the boulders apart with a force whose magnitude is equal to their gravitational attraction.  After this, we increase the magnitude of the pull by 10\% every 5 seconds.  The experiment continues this pull dynamics irrespective of whether breaking has occurred or not. Thus, once the bonds have broken the spheres will accelerate away from each other.

\subsection{Yield Strength of the Grain Matrix}

\begin{figure}[!h]
\begin{center}
\includegraphics[scale=0.4]{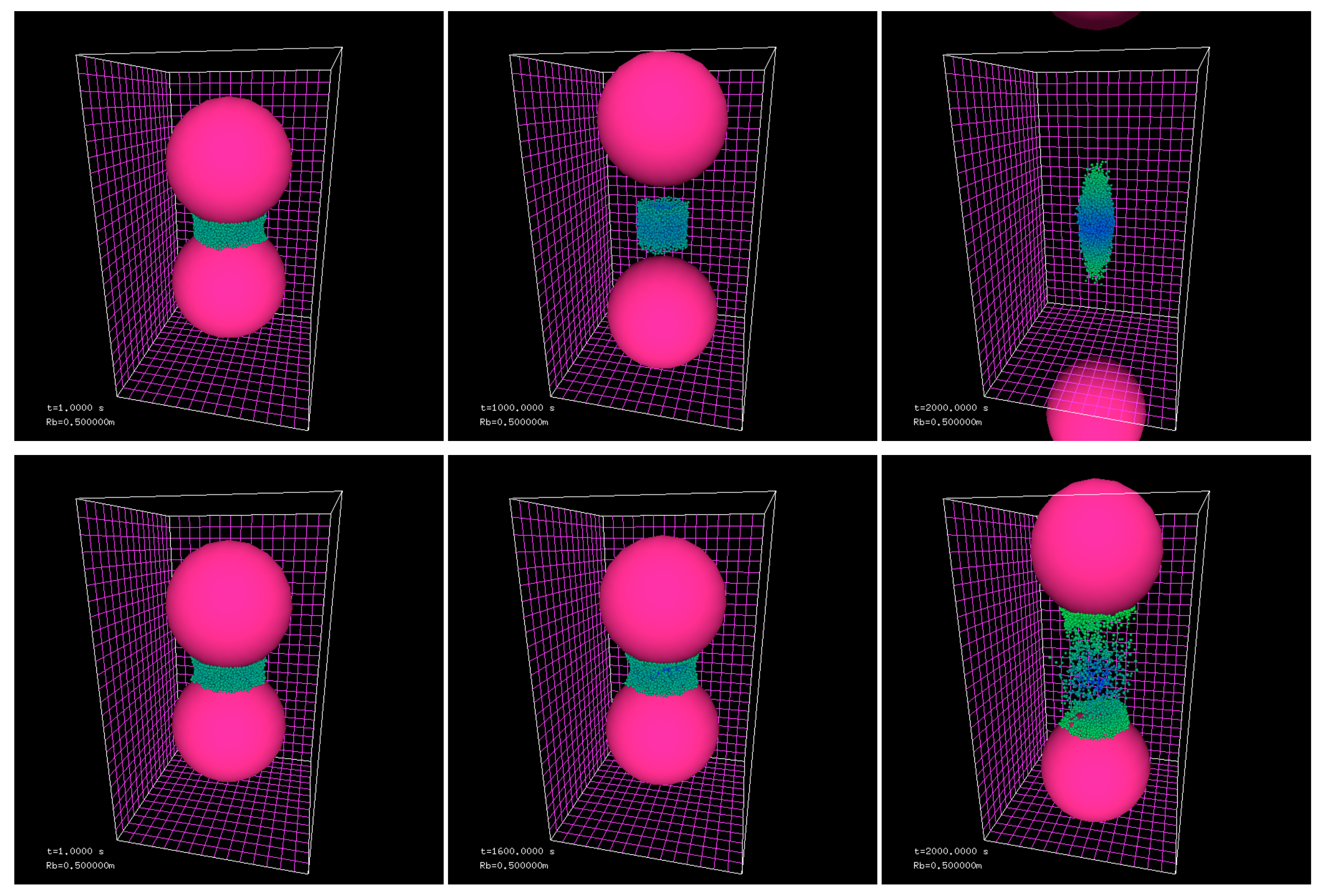}
\vspace{-12pt} 
\end{center}
\caption{\small L1$_{2-3}$ system with non-cohesive particles (top) and with cohesive particles (bottom).  From left to right t=0, 1000 and 2000 s.}
\label{snapshots}
\end{figure}

For a control experiment we simulated a L1$_{2-3}$ non-cohesive system and followed the pulling procedure as outlined above.  Snapshots of this simulation are presented in Fig.\ \ref{snapshots} (top).  Figure \ref{snapshots} (bottom) shows snapshots of the same L1$_{2-3}$ system, but with cohesive particles at t=0, 1000 and 2000 s.  Even though they initially have the same geometry, their dynamics are very different as the snapshots reflect.  The systems with more particles conserve a similar geometry to that shown in Fig.\ \ref{snapshots}, but with a larger bridge when more than one layer of particles is used.  The bridges that we formed have their maximum packing fraction near their centre, with values that range between 0.55 to 0.6.  The radii of the bridges increase with the number of particles, as expected, and are approximately 0.3, 0.35, 0.4 and 0.43 m for the 4 different layers.  All cohesive bridges remain virtually unchanged and deform only slightly from their initial shape during the simulations until they fracture.  This is not the case for the non-cohesive bridge, which in the absence of other forces conforms to the gravitational pull.

\begin{figure}[h!]
\centering
\includegraphics[scale=1]{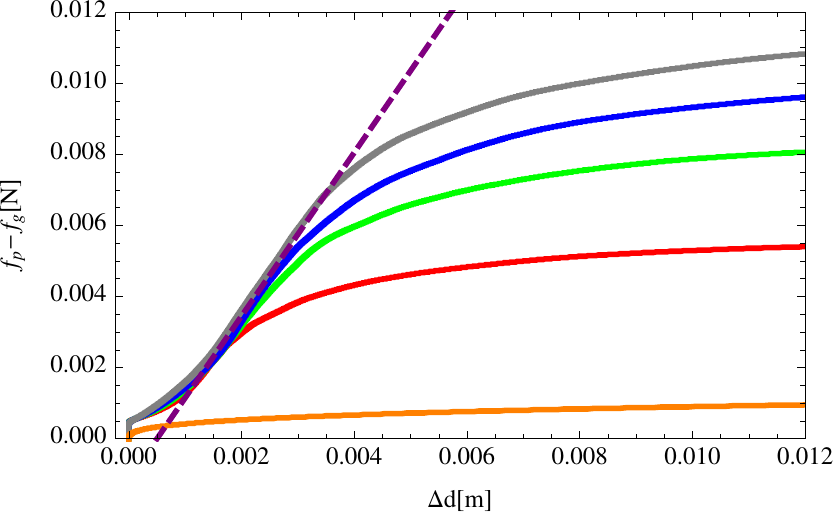}
\vspace{-12pt} 
\caption{\small Net pulling force over the top boulder vs. the variation of the distance between them for  cohesive (L1$_{2-3}$, L2$_{2-3}$, L3$_{2-3}$ and L4$_{2-3}$ are colored in red, green, blue, and grey respectively) and non-cohesive system (L1$_{2-3}$, orange).  The purple-dashed line is a guide to the eye; it has a slope of 2.3 N m$^{-1}$.}
\label{deformation-aver-complete}
\end{figure}

In order to determine the force needed to separate the two boulders in our systems we plot $\Delta f$, the difference between the pulling force ($f_p$) and the gravitational pull ($f_g$) due to the rest of the particles.  In Fig.\ \ref{deformation-aver-complete} the orange line represents the dynamics of the non-cohesive L1$_{2-3}$ system, whereas the other lines represent cohesive systems.  From this data it is evident that if cohesion is not present, the boulders begin to move apart as soon as a force greater than the gravitational attraction is exerted over them.  It is interesting to note that all the other systems start to strain when $\Delta f$ is about $4.5\times 10^{-4}$ N which is where the initial contacts between the regolith grains break.  After that, they all maintain a consistent slope of about 2.3 N m$^{-1}$ until the bridges are finally broken.  
The total magnitude of force required to ultimately break the regolith depends on the number of grains in the system.
The observed linear part of the dynamics in Fig.\ \ref{deformation-aver-complete} supports the overall goal of replacing the regolith with a soft potential that exhibits classical elastic and yield strength behavior.

\begin{figure}[h!]
\centering
\includegraphics[scale=0.33]{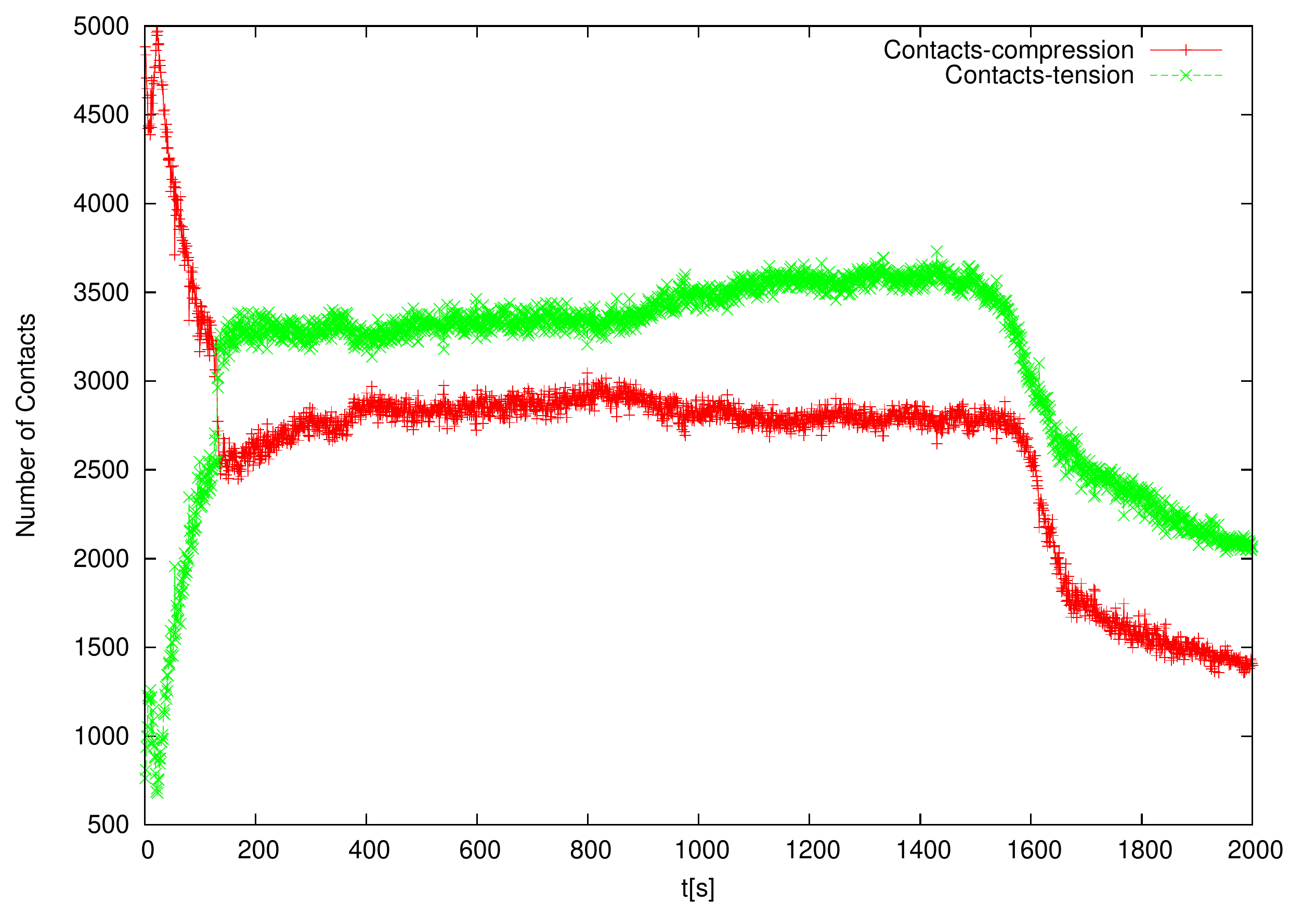}
\vspace{-12pt} 
\caption{\small Number of contacts in tension (green) and in compression (red) for an L1$_{2-3}$ system, as they evolve, during the duration of the simulation}
\label{contacts-L1}
\end{figure}

In order to obtain greater insight into the dynamics of the process it is necessary to understand why the boulders do not separate until the pull reaches a certain strength.  To do this, we need to realise that ultimately, it is the contacts between individual particles that control the macroscopic behaviour of the systems.  As the separation process starts there is a slight increase in the number of contacts, however the number of contacts that are in compression and in tension change drastically.  Fig.\ \ref{contacts-L1} shows the evolution of the number of contacts in tension (green) and in compression (red) for an L1$_{2-3}$ system, as they evolve, during the duration of the simulation.  From this figure it is observable that there is a significant change once the regolith breaks its initial contacts and engages the matrix of contacts more completely. After this point, the body begins to strain and we observe the elastic behavior in Fig.\ \ref{deformation-aver-complete}.  At this point the majority of contacts in the regolith switch to tension, essentially forming a series of chains through the regolith that become engaged and pulled into elongation under the increasing pull.

\begin{figure*}[ht!]
\centering
\includegraphics[scale=0.45]{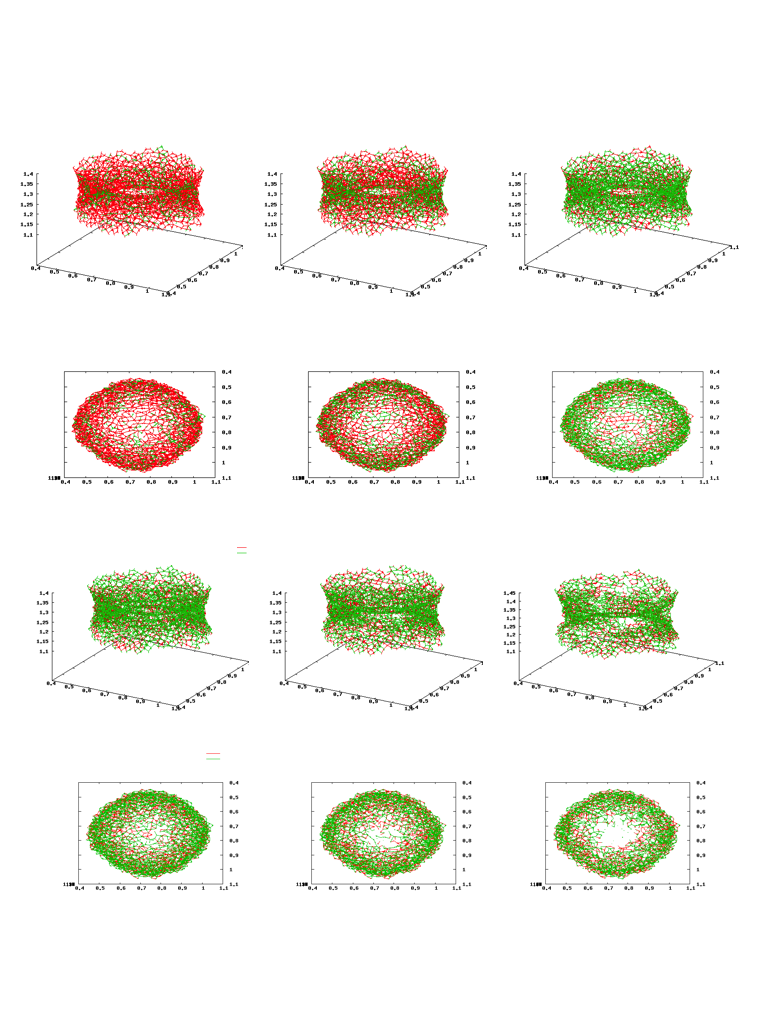}
\caption{\small Side and top views of the force chains in an L1$_{2-3}$ system during the pulling process. Red signifies a compressive chain and green signifies a tensile chain.}
\label{chains}
\end{figure*}

To explain the behavior of the bridges due to the dynamics of the contacts between the particles we look at the force chains that are being formed and how they behave.  Fig.\ \ref{chains} shows two views of the force chains in an L1$_{2-3}$ system during the pulling and breaking process.  The force chains in compression are marked in red, whereas those in tension are in green.  There is a clear difference between t=110s and t=120s, exactly at the point where the boulders begin to strain.  This and the results shown in Fig.\ \ref{contacts-L1} point to the appearance of the first fracture in the system which allows for a rearrangement of the particles and force chains.  From Fig.\ \ref{chains} we see that it is those particles near the center that break as they cannot re-accomodate to stretch the chains they form.  Once these central chains are broken the longer and more numerous chains away from the center become engaged and become tense.  This is why a stronger pull is needed to break a bridge with more particles.  Simply put, there are more chains that have to be broken before the bridge is completely fractured.

\begin{figure}[hb!]
\centering
\includegraphics[scale=0.45]{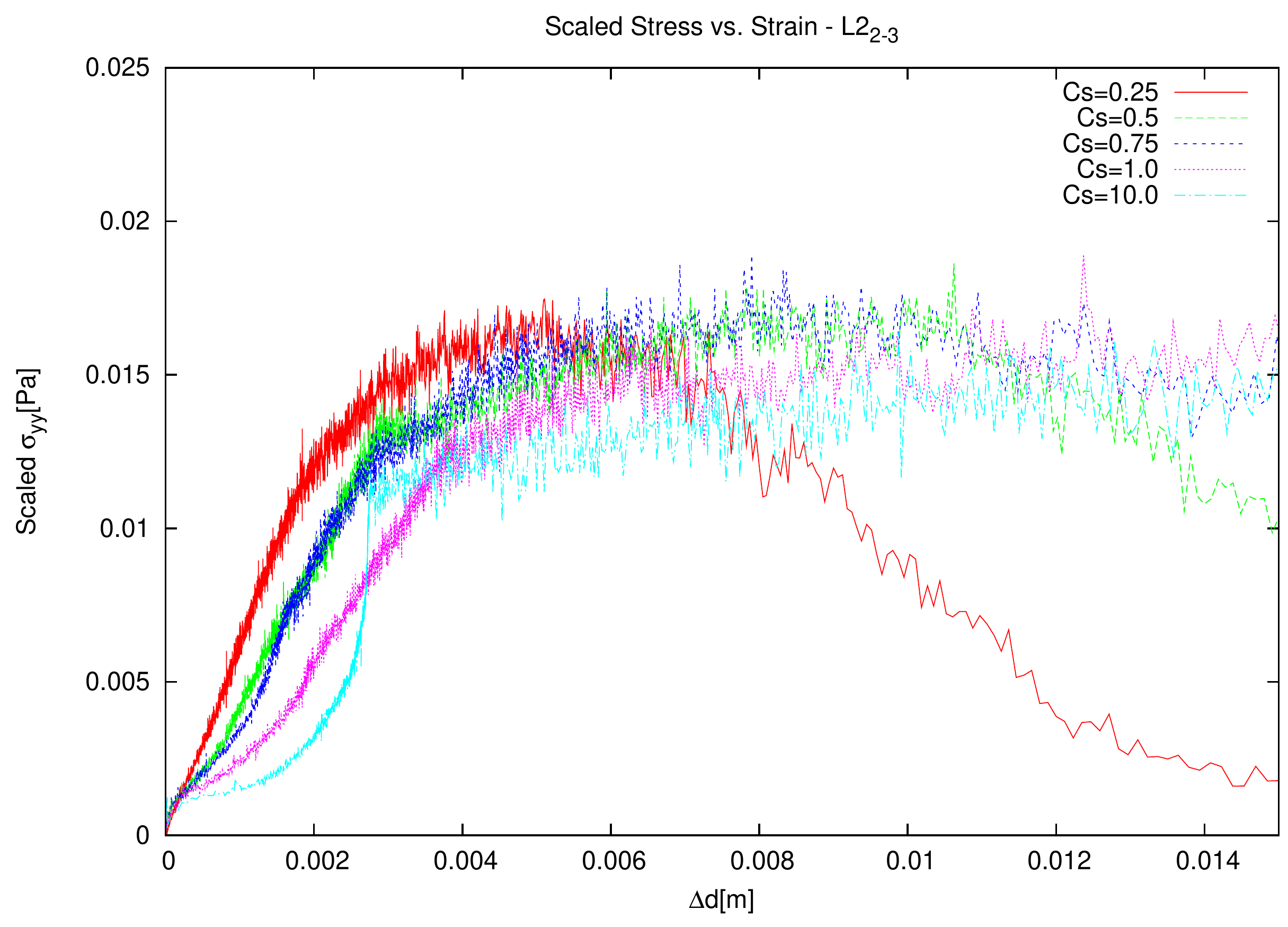}
\caption{\small Scaled normal stress of L2$_{2-3}$ systems, cohesive forces  have been scaled by factors of 0.25, 0.5, 0.75, 1.0 and 10.}
\label{scaled-stress}
\end{figure}

We note that the tension of 0.012 Pa required to break the bridges has the same value regardless of the number of particles used for a given grain size.  This is easily explained if we notice that an increase in the volume of the bridge, given by an increased number of particles, also increases the number of contacts linearly.  This leaves the tensile force per unit area needed to break the bridge unchanged.  Following the same train of thought, this stress should change with the strength of the cohesive forces involved.  To test this we have carried out simulations where we have scaled all the cohesive forces by factors ($C_s$) of 0.25, 0.5, 0.75 and 10.0; our scaled results are presented in Fig.\ \ref{scaled-stress}.  The scaling factor used for the normal stress is 1/$C_s$.  From these results, it follows that the tension needed to break a bridge for centimeter sized grains is $\approx 0.012\times C_s$Pa.  This marks the yield limit, defined as the end of the proportionality region in the stress-strain plot.  The strain of the bridge, on the other hand, seems to follow a more complicated scaling law, but in general it was never more than a few millimeters for our simulated systems.  The same behavior was observed in all our simulations and they all produced the same maximum normal stress before breaking.  Taking this approach, an increase in the cross-sectional area of the bridge increases the force needed to break it, yet keeps the yield stress the same. 

\subsection{Empirical Determination of Regolith Strength}

We note that the theory indicates that strength varies inversely with particle size, and thus we also test this relationship. However we must first develop an empirical correction to the ideal strength law derived earlier.  While Eq.\ \ref{ncontacts} gives us the total number of contacts across the cross-sectional surface, from Fig.\ \ref{contacts-L1} we can see that only about 60\% of the contacts are in tension and the same statistics should hold across this surface.  In addition, given that the particles are, in general, not aligned with the vertical, only a component of the tensile force contributes to the yield limit.  To correct for this we multiply the cohesive force by $\frac{3}{\pi}\sin(\pi/3)$, the average value of the function $\cos(x)$ in the interval [$-\pi/3,\pi/3$], the limits chosen to select those forces that are primarily in the given direction.  Moreover, our simulations reveal that the magnitudes of the forces between particles that are in tension are not centred around the value of the cohesive force of two average size particles.  In fact, the average of the this net tension is  
$\approx 11.25$ smaller than that.  The total tensile stress across the surface area for yield ($\sigma_{Y}$) would then be the result of multiplying our ideal strength by these additional correction factors, which results in
\begin{equation}
\sigma_{Y}=\frac{1.56\times10^{-4}}{\bar r_p}
\label{numeric-sigma}
\end{equation}

In the simulations shown in Fig.\ \ref{snapshots} the grains had an average radius of $\bar r_p=1.25\times 10^{-2}$ m, predicting a limiting $\sigma_{yy}=0.0125$ Pa. Simulations found a value of 0.012 Pa, showing excellent agreement with this calculation.
We note that this result is well over an order of magnitude weaker than the theoretical result we computed, assuming a packed, crystalline structure. This shows the effect of randomization in the regolith grain packing at the finest level.

\begin{figure}[hb!]
\centering
\includegraphics[scale=0.45]{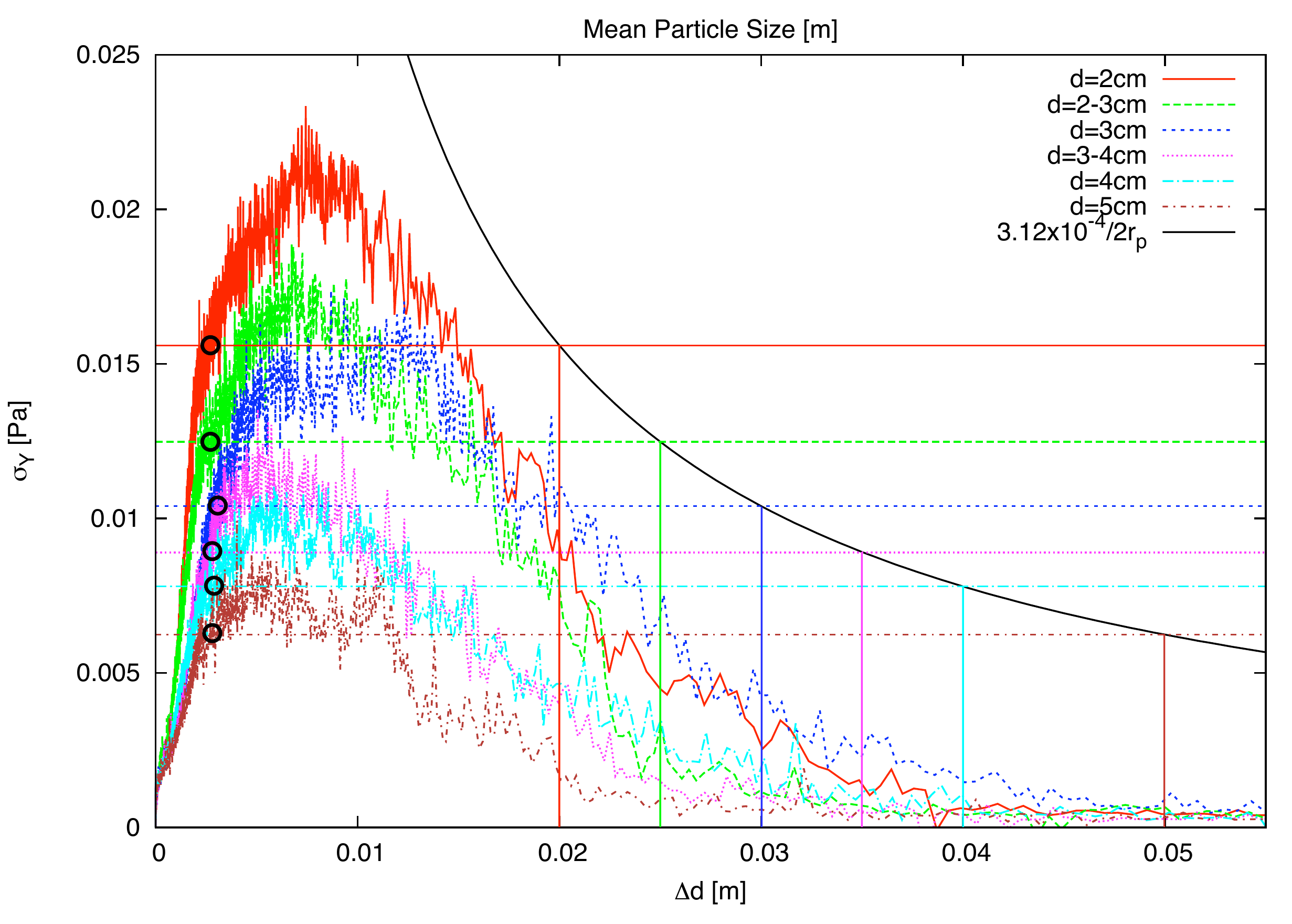}
\caption{Stress-strain relations for regolith of different size grains. The black line shows the predicted yield strength as a function of particle diameter -- the strain and the diameter sharing the same measurement tics.}
\label{fig:particle_size}
\end{figure}

The last equation shows the same $1/\bar r_p$ dependency of $\sigma_{Y}$ as seen in the ideal calculations, except now the appropriate grain size is the mean grain size. This was explicitly tested and all the simulations with monodisperse L1$_x$ systems and polydisperse L1$_{2-3}$ and L1$_{3-4}$ systems show the same numerical agreement, validating this equation.  Figure \ref{fig:particle_size} shows the calculated stress/strain curves for different particle sizes compared to the theoretical yield limit. The black line instead plots the predicted stress limit as a function of grain size. It must be noticed that in this analysis the size of the large boulders has no influence over the final outcome.  Therefore, in principle it should be applicable to any two boulders between which there are cohesive particles regardless of their size and even their shape.  Everything else being equal, the net cohesive force between two boulders depends on the narrowest cross-sectional area of the bridge perpendicular to the direction of the pulling force.

\subsection{A Model for Cohesion in a Rubble Pile}

An important application of this research is to simulations, as it provides a model for cohesion that is simple and scalable.  A first principles simulation that would attempt to capture the effect of fines on a macroscopic system containing boulders of order tens of meters would be intractable for a realistic $1/d^3$ size distribution. Our current findings of how interstitial-cohesive regolith behaves under tension shows that this simulation gap could be bridged by developing a soft cohesive potential that captures the strength we are specifying.  Specifically, the proportionality region in the stress-strain plots could be modeled with a linear spring with a strength limit, or in its simplest form, by a constant cohesive force at work only when boulders are in contact or within a given close proximity to each other.  This would allow researchers to implement the effect of cohesive regolith without the need to simulate each individual particle, saving significant time and resources.

\subsubsection{Testing the Model}

\begin{figure}[ht!]
\centering
\includegraphics[scale=0.4]{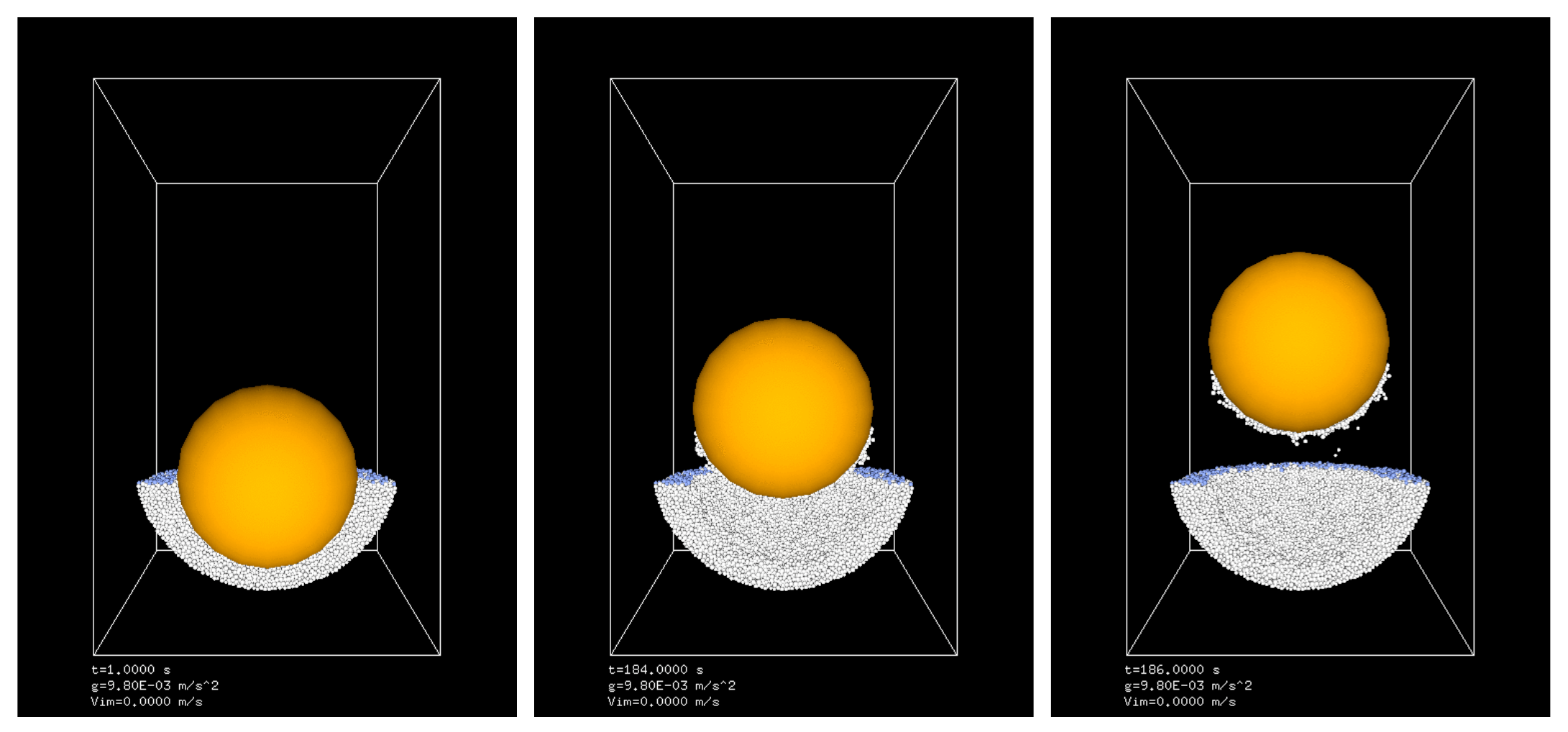}
\caption{A 1 m boulder being vertically pulled out  a granular pond; milli-g environment.}
\label{surf-coh}
\end{figure}

\begin{figure}[h]
\centering
\includegraphics[scale=0.33]{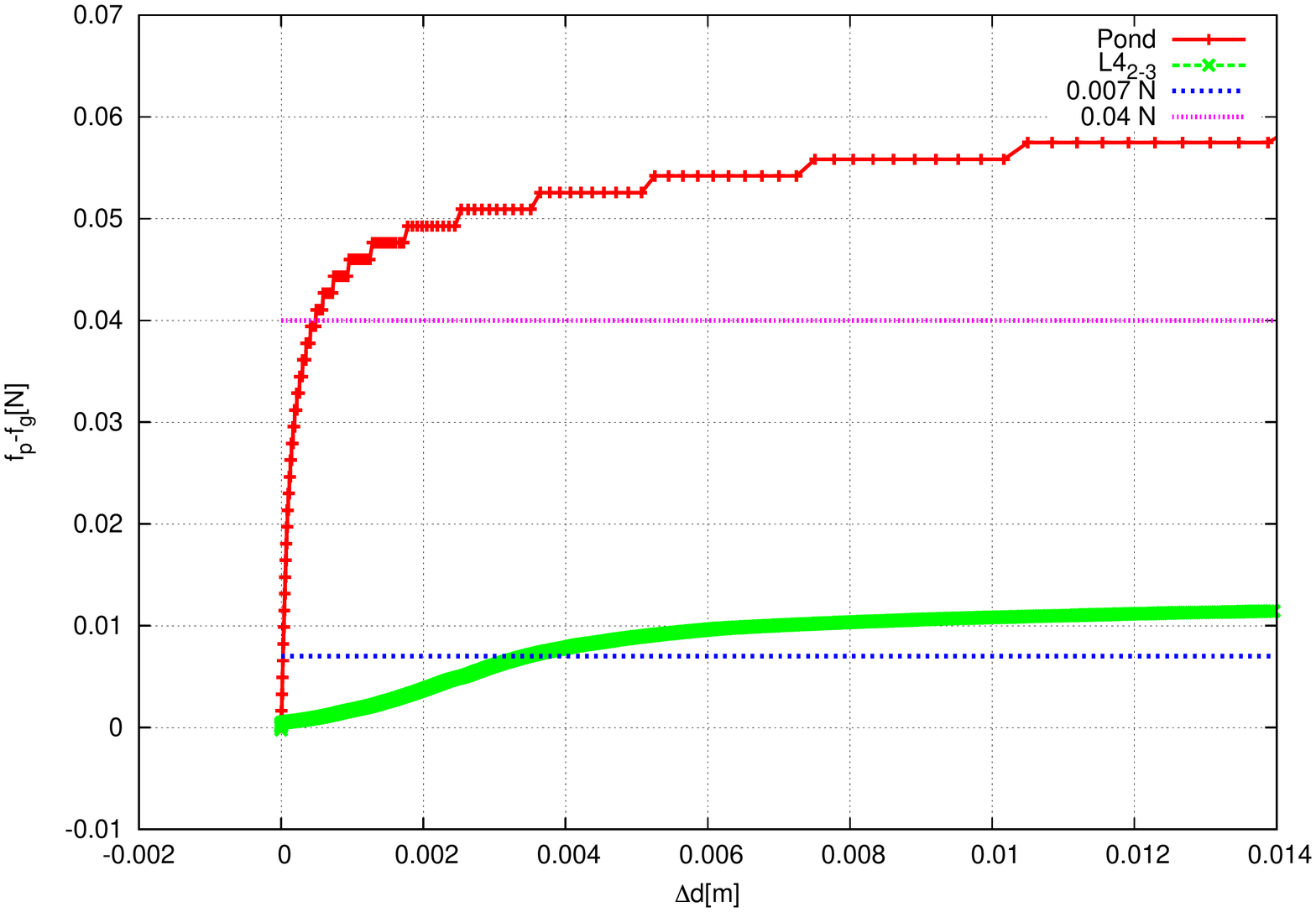}
\caption{Net pulling force over the boulder vs. its displacement.  Red: boulder in a pond, green: L4$_{2-3}$ system, the horizontal lines are only a guide to eye marking the force at which the boulder begins to move significantly.}
\label{all-comparison}
\end{figure}

To evaluate the effect on a boulder immersed in a pool of regolith, we simulated a 1 m boulder half-immersed in a hemispherical container filled with a polydisperse system with particles between 2 and 3 cm in size. 
Figure \ref{surf-coh} shows the simulation as the boulder is pulled out of the regolith.  Gravitational forces are calculated by means of an imposed gravitational field of 1 milli-g.  Figure \ref{all-comparison} shows that the magnitude of the force needed to detach the boulder from the regolith is about 0.04 N, a factor of 6 greater than the largest force needed to break an L4$_{2-3}$ system (0.007 N) with the same regolith size particles.  
Thus an increase in the contact area results in an increase in the net cohesive force. 

Based on this observation we develop a simple model that estimates the force needed to separate two boulders with interstitial regolith based on the projected area that links the two spheres. If the spheres are of equal size, we assume that the appropriate area over which the regolith chains will act equals the projected area of the spheres, or $\pi r^2$. As one sphere becomes larger, and in the limit approaches a flat plane, we assume that the effective area between the objects will become larger than the smaller sphere, representing the fact that regolith chains can form between the flat plane and the sphere beyond the projected area of the smaller sphere. A simple way to combine these areas into an effective area is to take their harmonic mean, or 
\beq
	A_{eff} & = & \frac{2}{\frac{1}{\pi R_1^2} + \frac{1}{\pi R_2^2}} 
\eeq
Then as $R_2 \gg R_1$ the effective area will be $2\pi R_1^2$, or twice the projected area of the single grain. This accounts for the ability of the regolith force chains to pull on the boulder from a slightly larger region than just the directly projected area.  
Then the cohesive force between two boulders in contact (or near contact) can be represented as the effective area multiplied by the yield strength of the regolith:
\begin{eqnarray}
F_c & = & \frac{2\pi R_1^2 R_2^2}{R_1^2+R_2^2} \frac{s_{yy}}{\bar r_p}
\label{net-cohesion}
\end{eqnarray}
where $F_c$ is the net cohesive force. For our earlier simulations we note that the interstitial area between the meter-sized grains is on the order of one-quarter less than their projected areas. Conversely, the simulation in this section mimics the more extreme extent, where the area from which regolith exerts a pull on the boulder is $2\pi R_1^2$. Combined together we note a predicted increase in strength on the order of 8, consistent with the observed increase in the strength of the bonds. 

\subsubsection{Simulation of a Rubble Pile with Cohesive-based Strength}

We now apply this model to a simulation to ascertain the effect of a regolith ``matrix'' on an NEO consisting of boulders. 
The motivation for these specific simulations is to show that the introduction of cohesion into a rubble pile can alter the failure behavior of that body for sufficiently strong cohesion. We note that our simulations begin to show the effect of cohesion for Bond numbers greater than 10, which is consistent with granular mechanics theories on when cohesive forces become important \cite{scheeres_cohesion}. 
We form a self-gravitating granular aggregate of 2000 boulders, 8-9 m in size following the procedure in \cite{sanchez_icarus}, forming an asteroid of size $\sim 100$ meters.  Taking the $s_{yy}$ from Eq.\ \ref{numeric-sigma} the resulting cohesive forces between boulders as a function of regolith grain size are shown in Table \ref{table-size}, where the Bond number $B$ is estimated by comparing the weight of the boulder that is furthest away from the centre of mass and on the surface of the asteroid to its cohesive force with its neighbor. Thus we see that we model the cohesive effect of regolith grains ranging from near meter size down to tens of microns.

\begin{table}[htbp]
  \centering
  \begin{tabular}{@{} ccc @{}}
    \hline
    B & $F_c [N]$ & $\bar r_p [mm]$  \\ 
    \hline
    0.001 & 0.0248 & 356.43  \\ 
    0.01 & 0.248 & 35.64 \\ 
    0.1 & 2.48 & 3.56 \\ 
    1 & 24.83 & 0.36 \\ 
    10 & 248.3 & 0.036 \\ 
    15& 372.5 & 0.024 \\ 
    20 & 496.7 & 0.018 \\ 
    \hline
  \end{tabular}
  \caption{Bond number, net cohesive force, and average radius of the regolith particles.}
  \label{table-size}
\end{table}

\begin{figure}
\centering
\includegraphics{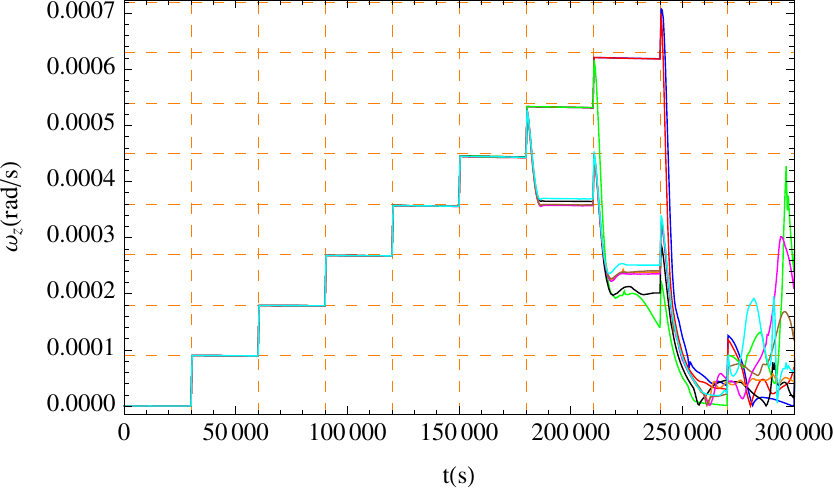}
\caption{Spin rate evolution of self-gravitating aggregates with different bond number for the particles on their surfaces.  The colors correspond to aggregates with bond numbers of of 20 (blue), 15 (red), 10 (green), 1 (black), 0.1 (orange), 0.01 (magenta),0.001 (brown), and 0 (cyan) for control.}
\label{spin-rate-coh2}
\end{figure}

Figure \ref{spin-rate-coh2} shows the spin evolution of aggregates formed by 2000 spherical boulders subjected to a spin rate that increases in steps (similar to the simulations described in \cite{sanchez_icarus}). These boulders have inter-particle friction and a cohesive force that acts as a constant between particles in contact.  The cohesive force is measured in terms of an effective Bond number for this asteroid, as discussed above.  The colors correspond to aggregates with bond numbers of of 20 (blue), 15 (red), 10 (green), 1 (black), 0.1 (orange), 0.01 (magenta),0.001 (brown), and 0 (cyan) for control.  The only systems that show a major difference with the others are the ones with B$\geq$10; the others reshape and disrupt at a similar spin rate. For the bond numbers greater than 10 the rubble pile fails in tension and does not undergo reshaping. Stronger regolith, due to smaller average grain sizes, will fail at progressively larger rotation rates.  Failure for these more cohesive systems occurs through fracture of the matrix material, as seen in our force chain computations. These numerical runs motivate the simpler analytical expressions we derive in the following for the limiting spin rate of a rubble pile. 

\section{Models for the Failure of a Rubble Pile}

We wish to apply the empirically computed yield strength of our regolith model to a simple model for the failure of a rubble pile asteroid strengthened by cohesive forces. To carry this out we will use the analytically simple Drucker-Prager failure criterion in combination with an analytical model for the interior stresses within a body due to the forces and loads applied to it. This model is intended to expose the basic relationships between a body's size, spin rate, strength and density and where it might be expected to conservatively fail.

In the following we review the known closed form stress fields for a rotating, self-gravitating ellipsoid and evaluate it following the Drucker-Prager failure criterion, inserting our determined cohesive strength. 

\subsection{Internal Stress Field}

For a simple model of a constant density ellipsoidal asteroid with no residual stress, the internal stress state due to gravitation and rotation can be computed.
The principal stresses within an ellipsoidal, gravitating body are stated in Holsapple  \cite{holsapple_original} as
\beq
	\sigma_x & = & - \rho/2 \left( \omega_\alpha^2 - \omega^2\right) \alpha^2 {\cal E} \\
	\sigma_y & = & - \rho/2 \left( \omega^2_\beta - \omega^2\right) \beta^2 {\cal E} \\
	\sigma_z & = & - \rho/2 \left( \omega^2_\gamma\right) \gamma^2 {\cal E} 
\eeq
where $\alpha \ge \beta \ge \gamma$ are the ellipsoid semi-major axes, $\rho$ is the constant density, 
\beq
	\omega^2_\alpha & = & 2\pi {\cal G} \rho \alpha\beta\gamma \int_0^\infty \frac{dv}{(\alpha^2 + v)\sqrt{(\alpha^2 + v)(\beta^2 + v)(\gamma^2 + v)}}
\eeq
represents the gravitational contribution from the ellipsoidal mass distribution (with appropriate permutations for $\omega^2_\beta$ and $\omega^2_\gamma$), $\omega$ is the constant spin rate about the $\gamma$ axis, and ${\cal E} = 1 - (x/\alpha)^2  - (y/\beta)^2  - (z/\gamma)^2$ defines how the stresses vary across the interior and is identically zero at the surface of the body. This is for an elastic model and neglects the possibly important residual stresses that can be built up as a body goes through plastic deformation. Such a detailed model is beyond the current work, as we are just looking for a clear representation of the important gravitational, inertial and cohesive forces at play within a rubble pile. 

We note that the quantity $\omega_\alpha$ is the spin rate at which the gravitational attraction equals the centripetal acceleration throughout that body's axis. For an elongate body we note that $\omega_{\alpha} < \omega_{\beta} < \omega_{\gamma}$, leading to $\sigma_x > \sigma_y > \sigma_z$ for a non-zero spin rate $\omega$. Conversely, for an oblate body we find that $\omega_{\alpha} = \omega_{\beta} < \omega_{\gamma}$, leading to $\sigma_x = \sigma_y > \sigma_z$ again for non-zero $\omega$. 

\subsection{Failure Criterion}

The Drucker-Prager failure criterion has been used to describe when a rubble pile will undergo deformation and failure \cite{holsapple_original, sanchez_icarus, sharma_DP}. This criterion incorporates the von Mises stress, the internal pressure, and the angle of friction between materials. We will use this failure criterion to derive a conservative limit on spin rate for failure, given a yield strength for the constituent material within the rubble pile. 

The von Mises stress is computed from the stress field as 
\beq
	\sqrt{J_2} & = & \sqrt{ (\sigma_x-\sigma_y)^2 + (\sigma_y-\sigma_z)^2 + (\sigma_z-\sigma_x)^2} / \sqrt{6}
\eeq
Given this, the Drucker-Prager stability criterion (for the tensile upper limit) is stated as
\beq
	\sqrt{J_2} & \le & k - \frac{2\sin\phi}{\sqrt{3}(3-\sin\phi)} \left( \sigma_x + \sigma_y + \sigma_z\right)
\eeq
where $\phi$ is the internal friction angle of the material and $k$ is a material constant related to cohesion. We use a simple relation between uniaxial yield strength and this parameter \cite{desai}, $k = \sigma_Y/\sqrt{3}$, where $\sigma_Y$ is the previously discussed yield strength. For a given yield strength, density and asteroid shape, this stability criterion identifies the failure envelope beyond which the asteroid undergoes plastic deformation. 

The Drucker-Prager criterion contains information on the body size, density, spin rate and body shape. It has been studied in detail in \cite{holsapple_spinlimits, sharma_DP} for a range of shapes, spin rates, and internal friction angles. S\'anchez and Scheeres \cite{sanchez_icarus} were able to show that deformation of a simulated rubble pile asteroid occurs when this criterion is violated. For the current application we just consider a few simplified cases of this law, enabling us to apply it generically to the asteroid population.

To motivate this we consider two separate cases where we have our spin rate $\omega > \omega_\alpha$, meaning that the body has tension along the longest axis, or along the equator for an oblate body. 

\paragraph{Elongate Body:} 
For an elongate body with $\alpha \gg \beta > \gamma$ we assume that $\sigma_x \gg \sigma_y, \sigma_z$, giving $\sqrt{J_2} \sim \sigma_x / \sqrt{3}$ and yielding the stability criterion
\beq
	\frac{\sigma_x}{\sqrt{3}} & \le & \frac{\sigma_Y}{\sqrt{3}} - \frac{2\sin\phi}{\sqrt{3}(3-\sin\phi)} \sigma_x
\eeq
Substituting $\sigma_x = - \frac{\rho}{2} \left(\omega_\alpha^2 - \omega^2\right) \alpha^2{\cal E}$ and solving for the spin rate yields
\beq
	\omega^2 & \le & \omega_\alpha^2 + \frac{2}{\rho \alpha^2 {\cal E}} \left( \frac{3-\sin\phi}{3+\sin\phi} \right) \sigma_Y 
\eeq
We note that the weakest point of the body occurs at its midpoint, under our idealizations, and thus take ${\cal E} = 1$. 

\paragraph{Oblate Body:} 
For an oblate body we assume that $\sigma_x = \sigma_y \gg \sigma_z$, giving $\sqrt{J_2} \sim \sigma_x / \sqrt{3}$ again but now yielding the stability criterion
\beq
	\frac{\sigma_x}{\sqrt{3}} & \le & \frac{\sigma_Y}{\sqrt{3}} - \frac{4\sin\phi}{\sqrt{3}(3-\sin\phi)} \sigma_x
\eeq
Substituting $\sigma_x = - \frac{\rho}{2} \left(\omega_\alpha^2 - \omega^2\right) \alpha^2{\cal E}$ and solving for the spin rate yields
\beq
	\omega^2 & \le & \omega_\alpha^2 + \frac{2}{3 \rho \alpha^2 } \left( \frac{3-\sin\phi}{1+\sin\phi} \right) \sigma_Y 
\eeq
where again we take ${\cal E} = 1$. We note that the gravitational spin rates $\omega_\alpha$ will be different between these two cases, however we are more focused on the strength of these models when $\omega^2 - \omega_\alpha^2 > 0$, independent of the precise spin rate at which the body goes into tension. 

\subsection{A Simple Model for Asteroid Failure}

We now consider both of these extreme models above. Comparing the cohesive terms only, we see that the ratio of the elongate over the oblate component is $3(1+\sin\phi) / (3+\sin\phi)$. Across all possible friction angles this ratio goes from $1 \rightarrow 1.5$ as $\phi$ goes from $0\rightarrow 90^\circ$. Thus, as a body spins beyond the its gravitational limit, the cohesive strength term is not a strong function of the assumed shape. 

Thus, due to its slightly simpler form, we use the elongate case as our simple model. 
\beq
	\omega^2 & \le & \omega_\alpha^2 + \frac{2}{\rho \alpha^2 } \left( \frac{3-\sin\phi}{3+\sin\phi} \right) \sigma_Y 
\eeq
Furthermore, as we wish to develop a conservative limit we note that the strength will be greatest for $\phi\rightarrow 90^\circ$, although only by a factor of 2 over the other extreme at $\phi = 0^\circ$. Inserting this we get a conservative limiting case for the spin rate as
\beq
	\omega^2 & \le & \omega_\alpha^2 + \frac{\sigma_Y }{\rho \alpha^2 } 
\eeq
We note that this relation has the proper characteristics as explored in more detail by Holsapple in \cite{holsapple_spinlimits}. Namely, for a given cohesive strength, the spin rate to disrupt them increases for smaller or less dense bodies. 

\begin{figure}[h!]
\begin{center}
\hspace*{-2.5cm}\includegraphics[scale=0.5]{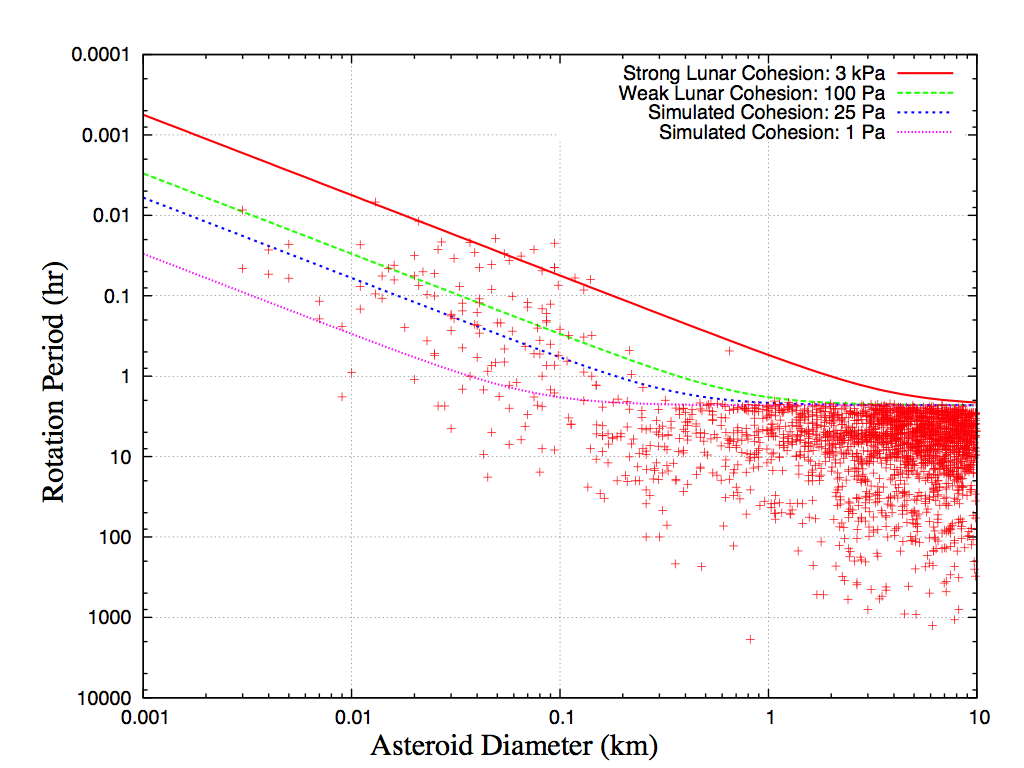}
\caption{Size/spin rate distribution. Lines show the theoretical spin limit for different assumed values of cohesion, computed using a conservative $90^\circ$ friction angle, yielding a larger spin-rate for a give quoted strength value.}
\label{fig:DP_all}
\end{center}
\end{figure}

This limit will be used to constrain possible strength values for small asteroids spinning beyond their gravitational rate. In Fig.\ \ref{fig:DP_all} we show the asteroid spin / size data overlaid with lines of theoretical spin limits for different levels of cohesive strength. This figure serves as motivation for our following discussion. 

\section{Observational Evidence for Cohesion}

There are a few aspects of the asteroid spin / size data that can be used to extract constraints on the level of cohesion which may be present in rubble pile asteroids. While none of these are definitive, we can show consistency with different aspects of this data set and an inferred level of cohesive strength. Perhaps the most compelling piece of evidence for cohesion among rubble pile bodies can be inferred from asteroid 2008 TC3 (which became the Almahata-Sitta meteorite fall) and the size distribution of binary asteroids. 

We note that the absolute values of cohesive strength quoted in the following are derived for a number of assumptions in our failure model. The assumption of an elongate body is conservative by at most a factor of 0.67 compared to an oblate body. The other is the assumption that the friction angle is taken as 90$^\circ$ to provide a bound on the strength. If the friction angle is reduced to 45$^\circ$ or $30^\circ$ the quoted strengths would decrease by a factor of 0.8 and 0.7, respectively. Combined, there is a conservative factor applied to the failure model that decreases the strength for failure by a factor of 0.47, meaning that the strengths quoted in Fig.\ \ref{fig:DP_all} could be a factor of $\sim 2$ larger. 

Finally, we also note that our empirically derived model presented in Eqn.\ \ref{numeric-sigma} for the relation between cohesion and mean grain size can be used to provide predictions and interpretations. Using this model a grain size of 1 micron ($\bar{r}_p = 0.5$ microns) predicts a strength of $\sim 300$ Pa, a grain size of 10 microns predicts a strength of 30 Pa, and a grain size of 100 microns a strength of 3 Pa. We will see in the following that the observed strength is consistent with a $\sim 25$ Pa level of strength, which then fits with the observations from Jewitt et al., the Itokawa sample return, and the lunar regolith data. 

\subsection{Binary Asteroid Size Cut-off}

Binary asteroids have a statistically significant cut-off at small sizes less than a few hundred meters \cite{margot}, commensurate with the initial increase in the asteroid maximum spin period. Failure of cohesive rubble pile asteroids through fission is consistent with a lower limit on the creation of binaries. If the failure spin limit is faster than local escape speed, then when fission occurs the bodies will immediately enter a hyperbolic orbit and escape from each other, providing no chance to dynamically interact and become stabilized. Such a period of orbital interaction is at the core of all competing, hypothesized binary formation mechanisms, such as are described in \cite{jacobson_icarus, walsh_nature}.  

\begin{figure}[h!]
\begin{center}
\hspace*{-2.5cm}\includegraphics[scale=0.5]{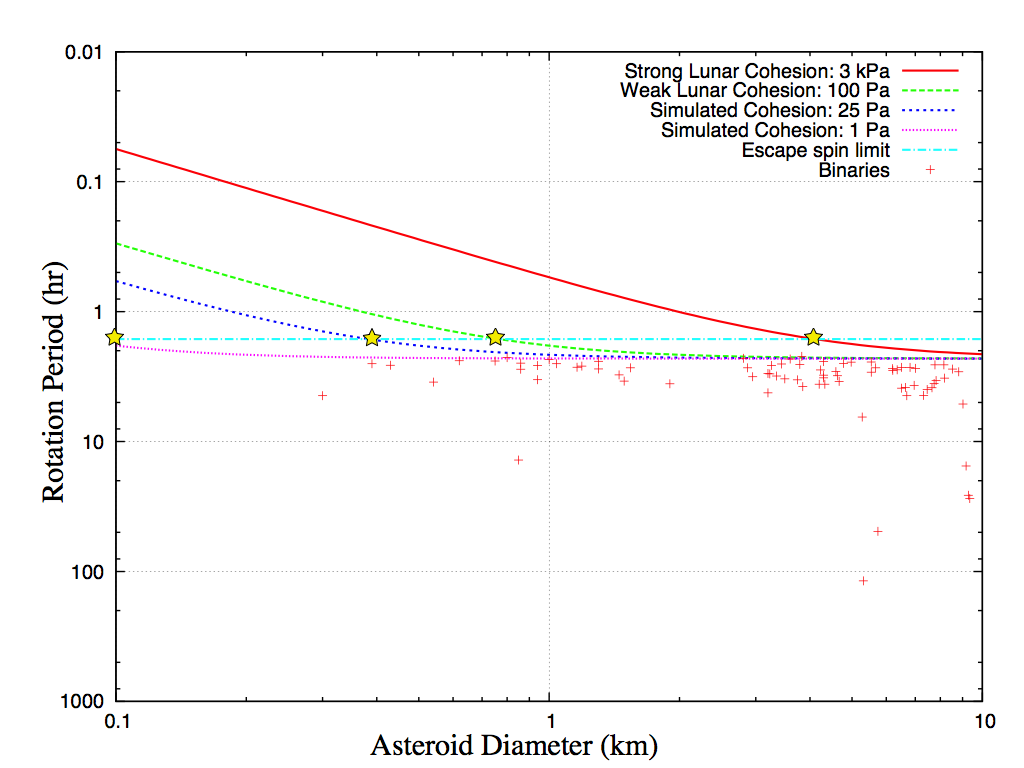}
\caption{Size/spin rate distribution of binary asteroids. Lines show theoretical strongest and weakest strengths considered. Only confirmed primaries with a data quality of 3 are plotted. Also shown is the limiting spin rate at which fissioned bodies would immediately escape from each other.}
\label{fig:binaries}
\end{center}
\end{figure}

In Fig. \ref{fig:binaries} we plot the limits for this maximum spin rate for a bulk density of 2.1 g/cm$^3$, delineating where fission leads to immediate escape. We note that the strong regolith predicts a binary cut-off at a much larger size, on the order of 4 km and the weak lunar regolith limit predicts a cutoff at 800 meters.
A cohesion of 25 Pa is consistent with the end of observed binaries at a size of 400 meters, with only one binary (from the highest quality rotation set) at a smaller size. At an even lower level of strength of 1 Pa we note that the cut-off should be at 100 meters, well below the observed cutoff. Thus, the presence of small, but non-vanishing, cohesion is consistent with the observed statistically significant lower size of binary asteroids. This also constitutes a prediction that smaller binary asteroids formed by rotational fission are not expected to be found less than this, or a similar, strength cut-off.

\subsection{Fast-Spinning Tumblers}

\begin{figure}[h!]
\begin{center}
\hspace*{-2.5cm}\includegraphics[scale=0.5]{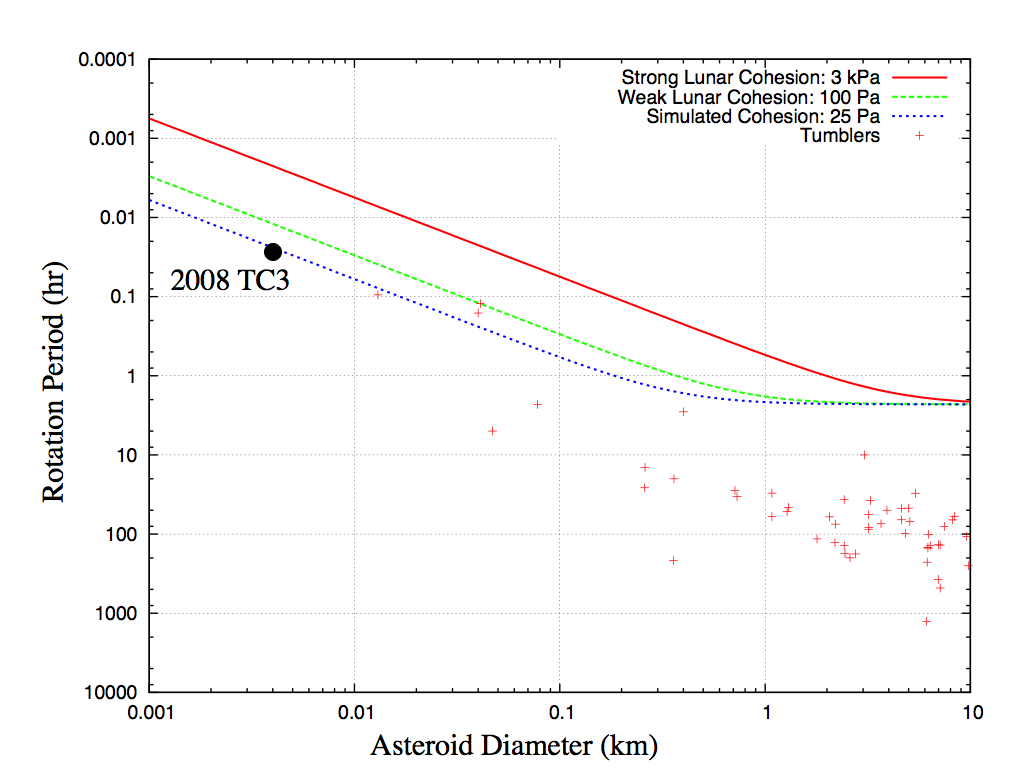}
\caption{Size/spin rate distribution of tumbling asteroids. Lines show theoretical strongest and weakest strength. The asteroid 2008 TC3 is the furthest body on the left. }
\label{fig:tumblers}
\end{center}
\end{figure}

There is another observable aspect of asteroid rotation for those bodies that fission at spin rates beyond the escape limit. In Scheeres et al.\ \cite{scheeres_cohesion} it was noted that a a rubble pile body which fissions across a surface of weakness when spinning faster than the escape limit will immediately enter a tumbling mode. This occurs as both portions of the body conserve their spin rate vector across the split but their mass distribution properties, i.e., principal moments of inertia, change abruptly as does their total angular momentum (although the sum is still conserved across both of the bodies). Assuming that the body was initially spinning about its maximum moment of inertia, the new bodies will instead commence to tumble due to the mismatch between principal moments of inertia and the spin vector. Unfortunately, the statistics on small, tumbling fast rotators is not very complete due to difficulties associated with reliably identifying this state \cite{pravec_tumbling}. Hence, there are only 6 such bodies in the current population with size below 100 meters. These, along with all other tumblers less than 10 km, are plotted in Fig.\ \ref{fig:tumblers}. We note that the rapid tumblers are all in close proximity to the weak regolith strength limit, consistent with the cited theory of failure. Conversely, this limit also provides a relevant prediction for where tumblers could be found, potentially serving as a test of the theory if the observed population of tumblers increases. Where such tumblers lie relative to the strength models also could provide insight into the strength of rubble pile asteroids at small sizes. 

Other explanations for rapid tumblers have not been clearly given in the literature, where the focus has been more on the dynamics of slow tumblers \cite{pravec_tumbling, vokrouhlicky_tumbling}. Applying the classical model for asteroid spin state relaxation to the small, fast tumblers \cite{harris_relax} yields relaxation times on the order of 1 Million years (see Fig.\ \ref{spinsize_harris}), which is shorter than the lifetime of an NEO and is not inconsistent with observing small bodies in this state. We note, however, that the physics of dissipation within small bodies is a topic that has not been explored. A dynamical explanation for the onset of tumbling with a body undergoing a net increase in spin rate has not been explicitly developed in the literature to date. Alternate possibilities include that the body was initially in a complex rotation state and subsequently spun-up, or that something in the spin-up process excited non-uniform rotation. Neither of these would preferentially predict that the body would reside close to the strength limits we have defined herein, however. 

\subsection{Failure Spin Rates}

Now we consider what the predicted spin / size distribution is as a function of cohesion. In Holsapple 2007 \cite{holsapple_spinlimits} it was shown that a cohesive strength of 10 kPa was needed to blanket all observed rapidly rotating bodies. However, as is seen for our 3 kPa limit in Fig.\ \ref{fig:DP_all}, this would predict rubble pile bodies would be able to fill up the area beneath the strength curve, which is definitely not seen at the mid-sized asteroids between 100 meters and 10 km. At 1 kPa this gap exists up to 1 km sized asteroids, and for 25 or 1 Pa it seems to disappear. This is not a strong constraint, but does indicate that the cohesive strength within rubble piles, if it exists, is small. 

The prediction from this model is that those asteroids spinning faster than the cohesive strength limit are ``competent'' or ``monolithic'' bodies, as a collection of grains and boulders would not have sufficient strength to hold together even with our ``van der Waals'' cement. The surfaces of these bodies can still have a regolith, however, comprised of grains ranging up to millimeter sizes or larger \cite{scheeres_cohesion}. The existence of such rapidly rotating boulders is entirely consistent with the supposition that asteroid rubble piles consist of a size distribution of grains. 

There are a few of these rapidly spinning monoliths of size greater than or equal to 200 meters, but the remaining members range from 100 meters down to the smallest sizes observed in the asteroid population. We note that the largest blocks observed on the asteroid Eros are of the order 100-150 meters \cite{robinson_eros}, potentially indicating this as the limiting size of competent boulders. The existence of such a maximum limit for a broken body is consistent with theories of material strength of bodies \cite{holsapple_spinlimits}. In this sense, the asteroids spinning faster than the weak regolith limit would represent the fractionation of rubble pile asteroids into their constituent pieces of bedrock, although many of these may still not have shed all particles from their surface. 

In contrast to those bodies spinning beyond our derived strength curves, asteroids below our strength limit could potentially represent rubble pile bodies. Of course, we might also expect some of these bodies to be monoliths as well, although discriminating between these types would require additional information on any particular body. Thus, the strength limits do not provide any strong constraints on which bodies spinning below our strength curves might actually be rubble piles. 

\subsection{Asteroid 2008 TC3}

As a specific case in point we can consider Asteroid 2008 TC3, which became the Almahata Sitta meteorite. This asteroid's unique observation history and size make it a compelling piece of evidence that is fully consistent with the model developed herein. This asteroid was observed prior to entering the Earth's atmosphere during the brief period after its discovery. Light curve observations of this body showed that it was spinning with a period of 100 seconds and also was in a tumbling rotation state \cite{TC3_rotation}. Its location is marked on Fig.\ \ref{fig:tumblers}, where it is seen to be the smallest known tumbler and only requires 25 Pa of cohesive strength to withstand disruption. Based on analysis of the meteorite fall, this body consisted of several different mineralogical types that constituted separate components in the parent asteroid. The implication is that the asteroid was a heterogenous mixture of rocks that had different mineralogies, i.e., that it could be described as a rubble pile \cite{jenniskens_TC3}. Analysis of the pre-entry observations and the meteorite falls also indicate that the body had significant macro porosity \cite{kohout_TC3}. The meteor was observed to break up high in the atmosphere, indicating a weak body and consistent with a weakly bound rubble pile \cite{borovicka_TC3, popova2011}. Furthermore, in \cite{borovicka_TC3} they note the presence of an abundance of micron-sized dust associated with the meteor, consistent with this dust comprising a ``substantial part'' of the total mass of the object. 

Thus, in this one asteroid we find several different elements of our theory. First, we note that the body is clearly a macroscopic composite of several different mineralogical types, consistent with the parent body being a collection of smaller, distinct components resting on each other. From the entry observations, it is also apparent that there was a substantial presence of finer dust grains associated with the asteroid. This speaks directly to our model of a rubble pile asteroid as being a consistent size distribution from larger to smaller grains. To date there does not appear to be any analysis of the possible size distribution that could be associated with this body, however. We also note that this rubble pile was spinning rapidly, but would have only required $\sim 25$ Pa of cohesive strength to bind the body together. This is consistent with the limits on strength that we have estimated based on direct modeling and through observations. In addition, the body was seen to be tumbling, which is also consistent with our model of what the spin state of a rubble pile asteroid should be following a fission event. Finally, the body was observed to fail very high in the atmosphere, indicative of an overall very low strength for the body. Thus we note that this body exhibits several different characteristics and elements that are consistent with our theory. While not a proof in any sense, it does show a consistency between our proposed model and an identifiable body that was thoroughly analyzed from several different aspects. 

\section{Discussion}

This paper lays out a theoretical model for how rubble pile asteroids may have some level of strength, simulates this process to better understand the predictions that it provides, develops a simple analytical model for spin limits on failure as a function of cohesion and size, and then analyzes the asteroid spin / size database for evidence of cohesion  amongst rubble piles. In this section we explore some of the predictions and implications of this work. 

The model we propose is that the finest regolith that exists in sufficient quantities to coat and connect the largest boulders within a rubble pile can supply cohesive strength to the body. This cohesive strength will allow the rubble pile to spin up to and beyond the point where centripetal accelerations exceed gravitational accelerations. From the simulations explored herein, we find that the overall strength of the body will equal the strength of the matrix corresponding to the average grain size in the distribution. This links the size distribution of grains within a rubble pile body to the strength of that body, potentially providing insight into the structure and distribution of grains within rubble pile bodies through more precise measurement or determination of the strength of rubble pile bodies and regolith in general. 
Alternately, experimental determination of size distributions that result from catastrophic impacts can also then be used to predict the strength of regolith and rubble pile bodies. 
The size distribution limit at which there will not be sufficient fines to serve as a matrix is an open question, but should be studied. We note that our simulation-derived strength model for regolith is consistent with our inferred cohesive strength and assumed size distribution of grains. It is also consistent with the published lunar data, if we restrict ourselves to the highest porosity upper layer. 

Specifically, we find that a reasonable value of observed cohesive strength of rubble pile asteroids is $\sim 25$ Pa. Using our currently derived model given in Eqn.\ \ref{numeric-sigma} we see that this corresponds to a mean particle sizes of 12.5 microns. Assuming that the binding regolith matrix arises from a $1/d^3$ size distribution, this indicates a minimum grain size of $\sim 8$ microns. This is surprisingly consistent with the Itokawa sample return \cite{tsuchiyama_science}, observations of disrupted body P/2013 P5 \cite{jewitt_P2013_A}, and our inferred size distribution from lunar regolith \cite{mitchell_lunar}. These convergent numbers provide another aspect of small bodies to be specifically tested with future sample returns and in-situ observations. 

While the current paper primarily relies on measurements of the S-Type asteroid Itokawa and properties of lunar regolith, different asteroid types and mineralogical properties are expected to result in different cohesive constants, different size distribution of shattered bodies, and potentially other characteristics that can influence the properties and strength of the matrix that serves as the van der Waals cement. As the population of asteroids with specific spectral asteroid type determinations increases, it may be possible to extract information about these parameters based on the observed spin / size distributions that these bodies have. Furthermore, future sample return missions to primitive asteroids such as the Hayabusa-2 and OSIRIS-REx missions will supply additional insight into the inferred strength and size distribution of material in these types of asteroids. 
Additionally, sorting asteroids for which we have spectral types into their own size / spin rate distributions could expose mineralogical and morphological differences between different classes of asteroids. This includes observations of binary sizes, tumbling asteroids, and spin rates as a function of size. Should systematic differences be observed in any of these distributions, it could be indicative of the different compositions and evolutionary paths that different asteroid types are subject to. 

The theory also makes several predictions on the observable consequences of cohesion in small asteroids. Primary is that asteroid fission rates should be a function of the overall size of the body, with smaller asteroids disrupting due to spin fission at faster rotation rates. With the recent observations of spin-disrupted asteroids \cite{jewitt_P2010_A, jewitt_P2013_A}, it seems that this could be further constrained by observations, and motivates photometric observations of those asteroids which have undergone such disruption to determine their current spin rate. Furthermore, if the primaries of these disrupted bodies are seen to be in a complex rotation state, this would also be consistent with the theory. 
In relation to these disrupted asteroids, it would be interesting to determine whether a larger secondary component may have been shed, leading to the current significant mass loss from the surface of the body. Whether or not such a secondary was formed would provide clear insight into the mechanics of asteroid fracture when spun to high rates. We note that this mode of fission is distinct from that observed in Pravec et al. \cite{pravec_fission}, as the fission mechanism consistent with those observations is linked to larger bodies that should behave more like a cohesionless system. Further, in those systems the signature of the primary spin rates as a function of secondary size is also consistent with a dissolution of the binary system after a significant, if relatively short (on the order of a year at most) period of mutual orbital evolution \cite{jacobson_icarus}. 

Beyond constraints on asteroid observations, this paper also provides an important and computationally efficient model for capturing the effect of a steep size distribution on the mechanics of a rubble pile body. By developing a specific model that captures the cohesive effect of fine grains interspersed between larger components, we enable modelers to bypass directly modeling steep size distributions down to small grain sizes. This can expand the number and range of simulations that are accessible to computational codes. 

\section{Conclusions}

This paper hypothesizes and evaluates the implications of a theory for the strength of rubble pile asteroids. The hypothesis is that the finest grains within an asteroid can serve as a ``cement,'' a cohesive matrix that binds larger boulders together into a body, allowing it to spin more rapidly than the surface disruption limit. The strength of such a matrix is shown to depend inversely on the mean grain size within the cohesive matrix. Implications of this strength limit are developed using stress theory and compared with the population of asteroids. We show consistency of our strength model with a cohesive strength in rubble pile asteroids on the order of 25 Pa. As input data for this model we consider observed limits in the asteroid spin/size data, including the binary cutoff, the strength envelope, and the presence of tumblers in the rapidly rotating population. This level of strength is also shown to be consistent with rubble piles having sufficient grains at the $\sim$ 10 micron size to connect larger boulders with each other. 

\section*{Acknowledgements}

P. S\'anchez acknowledges support from NASA's Planetary Geology and Geophysics program from grant NNX1OAJ66G. D.J. Scheeres acknowledges support from NASA's Planetary Geology and Geophysics program from grant NNX11AP24G. Both authors acknowledge support from NASA's Near Earth Object Observation program from grant NNXlOAG53G. The authors acknowledge discussions with Prof. M. Swift from the University of Nottingham which helped develop the original idea behind the granular mechanics simulations.
They also acknowledge useful reviews from the anonymous referees that helped improve this paper. 

\def\thesection{\Alph{section}}

 \setcounter{section}{0}

\section*{Appendices}

\section{Cubic Size Distribution}

The cumulative boulder size distribution for Itokawa has been measured to be of the form $N(r) = \frac{A}{r^3}$. Using this form for the cumulative distribution we can establish a number of useful results. 
Associated with this distribution is a maximum and minimum grain radius, $r_1$ and $r_0$, respectively. 
We interpret $N(r)$ to be the cumulative number of particles with radius between $r$ and the maximum size $r_1$. The term $A$ is initially chosen to agree with the observed number of largest boulders, $N_1$, such that $N(r_1)=N_1$. With this interpretation, we get the nominal form for the function:
\beq
	N(r) & = & N_1 \left(\frac{r_1}{r}\right)^3
\eeq

\paragraph{Size Frequency Density Function}

We interpret the cumulative distribution as the integral of a cumulative density function $n(r)$, defined as:
\beq
	N(r) & = & \int_{r}^{r_1} n(r) \ dr
\eeq
With this definition we can immediately note that $n(r) = - \frac{dN}{dr}$, leading to the cumulative density function
\beq
	n(r) & = & \frac{3N_1r_1^3}{r^4}
\eeq
We can also define a density distribution function that integrates to unity, denoted as $\bar{n}(r)$. We define this as
\beq
	\bar{n}(r) & = & \frac{n(r)}{\int_{r_0}^{r_1} n(r) \ dr}
\eeq
Carrying out this computation we find 
\beq
	\bar{n}(r) & = & \frac{3 r_1^3 r_0^3}{(r_1^3 - r_0^3) r^4} 
\eeq

\paragraph{Mean Grain Radius}

The mean grain radius is defined as
\beq
	\bar{r} & = & \int_{r_0}^{r_1} r \bar{n}(r) \ dr \\
	& = & \frac{3}{2} r_1 r_0 \frac{r_1+r_0}{r_1^2+r_1 r_0 + r_0^2} 
\eeq
Thus if $r_0 \ll r_1$ we find that $\bar{r} \sim \frac{3}{2} r_0$. 

\paragraph{Surface Area of Grains}

The total surface area in the rubble pile is computed as
\beq
	{SA}_T & = & \int_{r_0}^{r_1} 4\pi r^2 {n}(r) \ dr \\
	& = & 12\pi N_1 r_1^2 \left(\frac{r_1}{r_0} - 1\right)
\eeq
Thus we see that if $r_0 \ll r_1$, that we have an arbitrarily large total surface area.

A descriptive metric of surface area distribution is to find the particle radius $r_H$ at which the total surface area larger than this size equals the total surface area less than this size. This is found from
\beq
	\int_{r_0}^{r_H} 4\pi r^2 {n}(r) \ dr & = & \int_{r_H}^{r_1} 4\pi r^2 {n}(r) \ dr 
\eeq
It is easy to show that the value of $r_H$ that satisfies this equals
\beq
	r_H & = & \frac{2 r_0 r_1}{r_0 + r_1}
\eeq
which is the harmonic mean of the smallest and largest grains. Thus, if $r_0 \ll r_1$ we see that this equality occurs at $r_H \sim 2 r_0$. Thus we note that there is ``ample'' surface area for smaller grains to cover and come into contact with larger grains. 

More generally, the ratio of surface area from $r_0$ to a value $r_C$ over the surface area from $r_C$ to $r_1$ equals
\beq
	\frac{ \left( \frac{r_C}{r_0} - 1\right)} { \left( 1 - \frac{r_C}{r_1}\right)}
\eeq
For a system with $r_1 = 10$ meters, $r_0 = 1$ microns, and $r_C$ = 1 millimeter, we see that there is approximately 1000 times more surface area from microns to millimeters than from millimeters to decameters. 

\paragraph{ Volume of Grains}

The total volume of grains can be found by 
\beq
	{V}_T & = & \int_{r_0}^{r_1} \frac{4\pi}{3} r^3 {n}(r) \ dr \\
	& = & 4\pi N_1 r_1^3 \ln\left(\frac{r_1}{r_0}\right)
\eeq
and we note that it is dominated by the volume of the largest grains, as is expected.

Again it is interesting to find the radius $r_{half}$ such that the volume of grains lower than $r_{half}$ equals the volume larger than this radius. Solving for the total volume from $r_0$ to a size $r$ we find
$V_T(r) = 4\pi N_1 r_1^3 \ln (r/r_0)$. Equating this to $V_T(r_1) / 2$ and solving for $r$ gives us $r_{half}$. 
Setting the equation up and simplifying yields
\beq
	\ln (r_{half}/r_0) & = & \frac{1}{2} \ln (r_1/r_0) 
\eeq
Solving yields $r_{half} = \sqrt{ r_0 r_1}$, the geometric mean of the minimum and maximum grains.

\paragraph{Largest Grain}

Assume we have an asteroid with a mean radius $R$ and total volume $4\pi/3 \ R^3$. Equating this with the total volume of a distribution, we can find a relationship between the maximum and minimum boulder sizes and the total volume. 
Equating these volumes we find 
\beq
	\frac{1}{3 N_1} \left( \frac{R}{r_1} \right)^3 & = & \ln \left(\frac{r_1}{r_0} \right)
\eeq
Incorporating the packing fraction gives the result found in the paper.

\newpage 
\bibliographystyle{plain}
\bibliography{../../../bibliographies/biblio_article,../../../bibliographies/biblio_conferences,../../../bibliographies/biblio_misc,../../../bibliographies/biblio_books,../sanchez}

\end{document}